\documentclass[showpacs,showkeys,superscriptaddress,nofootinbib]{revtex4}
\usepackage{amssymb}
\usepackage{amsmath}
\usepackage{graphicx}
\usepackage{color}
\usepackage{hyperref}

\DeclareMathAccent{\ring}{\mathalpha}{operators}{"17}
\providecommand{\st}[1]{_{\text{#1}}}

\providecommand{\ut}[1]{^{\text{#1}}}

\def\onehalf{\frac{1}{2}}

\def\bra{\ensuremath{\langle}}
\def\ket{\ensuremath{\rangle}}

\def\ueq{\ut{eq}}

\def\tr{\mathrm{Tr}}
\def\Imat{\mathbb{I}}
\def\im{\mathrm{i}}

\def\cslb{\sigma_s}

\def\kv{\bv{k}}

\def\vv{\bv{v}}

\def\Fv{\bv{F}}
\def\bG{\bv{G}}
\def\bA{\bv{A}}
\def\bL{\bv{L}}

\def\cv{\bv{c}}

\def\rv{\bv{r}}
\def\b0{\bv{0}}
\def\uLB{\ut{LB}}
\def\uint{\ut{int}}

\newcommand{\mb}{\mathbf}

\newcommand{\bitem}{\begin{itemize}}
\newcommand{\eitem}{\end{itemize}}
\newcommand{\benum}{\begin{enumerate}}
\newcommand{\eenum}{\end{enumerate}}
\newcommand{\bblock}[1]{\begin{block}{#1}}
\newcommand{\eblock}{\end{block}}
\newcommand{\bmini}[1]{\begin{minipage}{#1}}
\newcommand{\emini}{\end{minipage}}
\newcommand{\btab}[1]{\begin{tabular}{#1}}
\newcommand{\etab}{\end{tabular}}
\newcommand{\btabn}[1]{\begin{tabular}{#1}}
\newcommand{\etabn}{\end{tabular}}
\newcommand{\beq}{\begin{equation}}
\newcommand{\eeq}{\end{equation}}
\newcommand{\beqn}{\begin{equation*}}
\newcommand{\eeqn}{\end{equation*}}
\newcommand{\bmult}{\begin{multline}}
\newcommand{\emult}{\end{multline}}
\newcommand{\bsplit}{\begin{split}}
\newcommand{\esplit}{\end{split}}

\newcommand{\bv}[1]{\mathbf{#1}}

\begin{document}

\title{Langevin theory of fluctuations in the discrete Boltzmann equation}

\author{M. Gross}
\affiliation{Interdisciplinary Centre for Advanced Materials Simulation (ICAMS), Ruhr-Universit\"at Bochum, Stiepeler Strasse 129, 44801 Bochum, Germany}
\author{M. E. Cates}
\affiliation{ SUPA, School of Physics, University of Edinburgh, James Clerk Maxwell Building, Kings Buildings, Edinburgh EH9 3JZ, UK}
\author{F. Varnik}
\affiliation{Interdisciplinary Centre for Advanced Materials Simulation (ICAMS), Ruhr-Universit\"at Bochum, Stiepeler Strasse 129, 44801 Bochum, Germany}
\affiliation{Max-Planck Institut f\"ur Eisenforschung, Max-Planck Str.~1, 40237 D\"usseldorf, Germany}
\author{R. Adhikari}
\affiliation{The Institute of Mathematical Sciences, CIT Campus, Tharamani, Chennai-600113, India}

\pacs{47.11.-j, 47.10.-g, 47.55.-t}
\keywords{Fluctuating Lattice Boltzmann equation, Non-ideal gas}

\begin{abstract}
The discrete Boltzmann equation for both the ideal and a non-ideal fluid is extended by adding Langevin noise terms in order to incorporate the effects of thermal fluctuations.
After casting the fluctuating discrete Boltzmann equation in a form appropriate to the Onsager-Machlup theory of linear fluctuations, the statistical properties of the noise are determined by invoking a fluctuation-dissipation theorem at the kinetic level.
By integrating the fluctuating discrete Boltzmann equation, a fluctuating lattice Boltzmann equation is obtained, which provides an efficient way to solve the equations of fluctuating hydrodynamics for ideal and non-ideal fluids.
Application of the framework to a generic force-based non-ideal fluid model leads to ideal gas-type thermal noise. Simulation results indicate proper thermalization of all degrees of freedom.
\end{abstract}

\maketitle

\section{Introduction}

Thermal fluctuations are important in a wide variety of mesoscopic flows in soft matter and biological physics. For example, thermal fluctuations are needed to produce diffusion in colloidal suspensions, to excite internal degrees of freedom in polymers and membranes suspended in a fluid, and to generate capillary waves on a fluid-fluid interface \cite{BarratHansen_book}. Even in a simple fluid, thermal fluctuations lead to subtle non-linear effects like long time tails and the renormalization of transport coefficients \cite{zwanzig_book}. Such effects are most pronounced near critical points where fluctuations dominate \cite{hohenberg_halperin}.

Theoretically, thermally fluctuating mesoscopic flows are most conveniently dealt with within the framework of fluctuating hydrodynamics \cite{Landau_FluidMech59, Reichl_book}. This approach, pioneered by Landau and Lifshitz for simple fluids, adds stochastic terms to the \emph{mechanical} equations describing the flow, thereby enabling its \emph{statistical mechanical} description. An important ingredient in this formulation is the fluctuation dissipation theorem (FDT) relating the variance of the stochastic terms to the Onsager coefficients of the fluid. In Landau and Lifshitz's original formulation, the FDT relates the variance of the random stresses to the viscosities of the fluid. Thus an FDT-respecting fluctuating hydrodynamic description allows one to calculate correlations in equilibrium as well as responses which deviate only weakly from equilibrium. This has led to the widespread application of the method to complex fluid flow \cite{sengers_book}.

Numerically, the solution of fluctuating hydrodynamic equations poses a serious challenge \cite{bell_methods_2007, donev_accuracy_2010}. Numerical solutions of the Navier-Stokes equations, particularly in complicated geometries, are difficult even in the absence of thermal fluctuations. This has led to a whole range of innovative numerical methods for the Navier-Stokes equations, of which the lattice Boltzmann method (LBM) stands out due to its simplicity and easily parallelizable nature. The basic version of the method is used to model isothermal, incompressible fluid flow, at both high and low Reynolds numbers \cite{succi_book}. Extensions of the basic method allow for flows of complex and multicomponent fluids with microstructure \cite{aidun_rev2010}.

The basic LBM ignores thermal fluctuations. Indeed, LB methods developed out of the need to eliminate thermal fluctuations inherent in their ancestor, the lattice gas method, which needed excessive averaging over the noise to extract results pertinent to mean fluid flow \cite{mcnamara_prl1988, higuera_succi_epl1989, higuera_epl1989}. Thus, the LBM as a preaveraged version of the lattice gas excluded all thermal fluctuations, making it inapplicable to the whole range of complex fluid flows mentioned above.
Thermal fluctuations were partially reinstated in the LBE in the seminal work of Ladd \cite{ladd_1994}, where, closely following the formulation of Landau and Lifshitz, fluctuations are added only to the transport degrees of freedom of the LBE, i.e.\ the stresses. However, besides the conserved moments of mass and momentum and the non-conserved transport moments representing the stresses in the fluid, there exist in any LBE model also higher-order degrees of freedom, the so-called ghosts. The ghost sector, which is coupled to the transport sector at all but the largest length scales, acts as a sink for the thermal stress fluctuations, and thereby destroys the balance between fluctuation and dissipation. Consequently, the thermalization of the fluid remains incomplete. This is reflected in a similarly incomplete thermalization of other degrees of freedom that may be coupled to the fluid, for example colloidal particles.
It was shown in \cite{adhikari_fluct_2005, duenweg_statmechLB_2007} how a fluctuating lattice Boltzmann equation (FLBE) can be constructed which offers a fully consistent discretization of the equations of fluctuating nonlinear hydrodynamics for an isothermal ideal fluid. This method reinstates the fluctuations in the ghost sector that were absent in previous work, thereby achieving a complete thermalization of the fluid at all length scales, not only the largest ones.
Notably, the general theory of LB-Langevin equations was also derived in an early account by \cite{dufty_lblangevin_1993}.

Recently, the Langevin approach of \cite{dufty_lblangevin_1993, adhikari_fluct_2005}, originally devised for the ideal gas, has been generalized to describe thermal fluctuations in the LBE for a class of non-ideal fluids \cite{gross_pre2010}. There, non-ideal fluid models based on a square-gradient (Ginzburg-Landau) free energy functionals were considered, implying that density fluctuations in such a fluid are spatially correlated with a correlation length which is proportional to the theoretical width of the diffuse liquid-vapor interface. This is in contrast to the ideal gas, where equilibrium correlations are generally absent for all degrees of freedom.
Clearly, the non-trivial structure factor of a non-ideal fluid has to be faithfully represented by the correlations of the LB population densities.
Using results of continuum kinetic theory, a general ansatz for these correlations has been proposed, noticing, however, that certain models might require modifications to account for implementation-specific details \cite{gross_pre2010}. This was, in particular, found to be the case for the model of Swift et al.\ \cite{swift_lattice_1995}, where, owing to the underlying modified equilibrium distribution, a spatially correlated form of thermal noise arises.

In the present paper, we show how thermal noise can be incorporated into both ideal and non-ideal fluid versions of the discrete-velocity Boltzmann equation (DBE).
The DBE is a precursor of the LBE that arises from the continuum Boltzmann equation by restricting velocity space to a finite number of velocities, while keeping position space and time continuous \cite{abe_jcompp1997, shanHe_DBE_1998}.
The fluctuating DBE (FDBE) is put forward here as a unifying starting point to construct fluctuating LB models and as a theoretical link between two successful frameworks of non-equilibrium physics: the LB method and the theory of linear regression of fluctuations due to Onsager and Machlup \cite{onsager_machlup_1953, deGroot_Mazur}.
The utility of the FDBE is illustrated by re-deriving the fluctuating LBEs of the ideal gas \cite{adhikari_fluct_2005} and the modified-equilibrium model \cite{gross_pre2010}.
Moreover, based on the model of He, Shan and Doolen \cite{he_dbe_1998}, we propose a FDBE for a generic force-based non-ideal fluid model and outline the requirements that have to be fulfilled in order to obtain a mathematically well-defined noise covariance.
It is shown that the noise of this model is of the same form as in the ideal gas, in agreement with the physical notion that reversible interactions, such as those present in this type of non-ideal fluid, should not contribute to dissipation. This derivation constitutes the central result of the present work.

Due to the presence of lattice corrections in the LBE expressions of the densities and fluxes and the generally non-linear form of the advection operator, the analysis of force-based non-ideal fluid models is technically difficult within the FLBE approach presented in \cite{gross_pre2010}. The FDBE constitutes in this case a preferable starting point for the analysis of thermal noise in the LBE.
A word of caution, however, is in order, as the DBE necessarily ignores also details of the spatial discretization scheme of the force, which can be an important ingredient in a LB model \cite{lee_fischer_2006}.
Only as long as the discretization scheme is conservative and does not give rise to any additional irreversible terms, can both LBE and associated DBE be regarded as equivalent concerning their behavior under thermal noise.
The present framework should thus be seen as complementary to the FLBE approach of \cite{gross_pre2010}.

To set our work in context, it must be mentioned that the study of fluctuations in the \emph{continuous} Boltzmann equation has a long history. The Boltzmann \emph{stosszahlansatz} effectively removes fluctuations from the Boltzmann equation, giving a mean-field description of the fluid. However, fluctuations can be restored by promoting the Boltzmann equation into a Langevin equation. This idea was first implemented by Kadomtsev \cite{kadomtsev_1957}, and followed by Abrikosov and Khalatnikov \cite{abrikosov_1958}. Bixon and Zwanzig \cite{bixon_zwanzig_1969} gave a heuristic derivation of the fluctuation-dissipation theorem for the fluctuating Boltzmann equation. The work of Fox and Uhlenbeck \cite{fox_uhlenbeck_1970a, fox_uhlenbeck_1970b} cast the theory into the framework developed by Onsager and Machlup for the linear regression of fluctuations, suitably generalised for the mixed parity of the Boltzmann equation under time.
Fox and Uhlenbeck paid adequate attention to the conservation laws of mass, momentum and energy when constructing the fluctuation terms and deriving the FDT relation for the variances. From this point of view, our work can be looked upon as a consistent discretization in velocity space of the Boltzmann-Langevin equation developed by the above authors and an extension thereof to include non-ideal fluid interactions. When combined with a numerical scheme for temporally integrating the discrete-velocity Boltzmann-Langevin equation, we obtain, among others, the FLBE previously derived in \cite{dufty_lblangevin_1993, adhikari_fluct_2005, gross_pre2010}.
Subsequent to the work of Fox and Uhlenbeck, Kac and Logan \cite{kac_logan_1976} showed that the fluctuating Boltzmann equation can be derived from a master equation description of fluctuations in phase space, due to Siegert \cite{siegert_1949}. The phenomenological assumption of an additive noise could then be rigorously justified. In a recent work, D\"unweg et al.\ \cite{duenweg_statmechLB_2007} have obtained the ideal gas FLBE of \cite{adhikari_fluct_2005} from a coarse-graining of a lattice-gas model. In this sense, their work is the analogue for the FLBE of the work of Kac and Logan for the fluctuating Boltzmann equation. Their work, based on deterministic equations for probability densities, is thus complementary to the approach to the ideal gas presented below, based on stochastic equations for the fluctuating variables.

The paper is organised as follows. In the next section, we summarise the Langevin theory of fluctuations for the Boltzmann equation and present key results of kinetic theory necessary for the extension of the theory to the non-ideal fluid. In section III we discuss the theory of finite-dimensional moment expansions in the discrete velocity-space, which facilitates the construction of the noise and the derivation of the FDT.
The main new contribution of the present work is discussed in section IV, where we introduce the fluctuating DBE for the ideal and non-ideal fluid models and provide expressions for the noise covariances. In section V, we show how the FDBE may be integrated in time to obtain the FLBE, providing an efficient numerical method for solving the equations of fluctuating hydrodynamics for ideal or non-ideal fluids. In section VI, simulations of a particular implementation of the force-based FLBE are performed for a number of basic test cases.
We conclude with a discussion of further extensions of the method and applications to problems of complex fluid flow.

\section{Fluctuations in the Boltzmann equation}
The phenomenological theory of fluctuations in the Boltzmann equation of a dilute gas, as we mentioned before, has been studied by several authors. The treatment which is most relevant for us is that of Fox and Uhlenbeck \cite{fox_uhlenbeck_1970b}. They show that the linearized Boltzmann equation with a Langevin noise term is of the form considered in the Onsager-Machlup linear theory of fluctuations \cite{onsager_machlup_1953, deGroot_Mazur}, as summarized in the next section.

\subsection{Linear theory of fluctuations}
The theory of linear regresssion of fluctuations, as originally proposed by Onsager and Machlup \cite{onsager_machlup_1953, deGroot_Mazur}, treats fluctuations of variables which are either even or odd under time-reversal symmetry. The Boltzmann equation has a mixed character, due to the presence of the reversible advective term and the irreversible collision term (see below). Fox and Uhlenbeck, therefore, generalise the Onsager-Machlup theory to such situations \cite{fox_uhlenbeck_1970a}. They consider fluctuations of a set of variables $\mathbf{a}(t) = [a_1(t), a_2(t), \ldots, a_N(t)]$ whose probability distribution at equilibrium is given by
 \beq
  P^{eq}(\mathbf{a}) = \frac{1}{Z} \exp \left(-\onehalf \mathbf{a} \cdot \mathbf{G}^{-1} \cdot \mathbf{a}^\dagger \right)
 \label{eq:steady_state_prob_dist}
 \eeq
where $Z$ is a normalization constant and $\mathbf{a}^\dagger$ is the conjugate transpose of $\mathbf{a}$.
The matrix of the equal-time correlations of the $a$'s is fixed by the entropy matrix $\mathbf{G}$ \footnote{In the original treatment of Fox and Uhlenbeck, the \emph{inverse} of $G$ is called `entropy matrix'},
\beq
G_{ij} = \bra a_i(t) \, a_j^*(t) \ket\,,
\eeq
where $^*$ denotes complex conjugation.
The $a$'s are taken to obey linear Langevin equations of the form
 \beq
  \frac{\partial}{\partial t} \, \mathbf{a}  = -\mathbf{L} \cdot \mathbf{a} + \mathbf{\xi}\,,
  \label{eq:lin-langevin}
 \eeq
where, $\mathbf{L}$ is a general time-evolution operator whose eigenvalues must have non-negative real parts. By convention, $\mb{L}$ is often written in terms of a symmetric and antisymmetric matrix, $\mathbf{L} = \mb{S} + \mb{A}$. Typically, the symmetric matrix $\mb{S}$ describes relaxation, and hence, dissipation, while the antisymmetric matrix $\mb{A}$ contains the reversible part of the dynamics (e.g.\ free streaming and thermodynamic interactions). The random noise term $\mb{\xi}(t)$ is a zero-mean Gaussian process, and is therefore specified entirely by its second moment.
It can be shown \cite{fox_uhlenbeck_1970a, zwanzig_book} that the Langevin equation generates trajectories whose one-point probability distribution converges to eq.~(\ref{eq:steady_state_prob_dist}) if and only if the noise variance satisfies a fluctuation-dissipation relation,
 \beq
  \langle \xi_i(t) \xi_j(t') \rangle = \left( \mathbf{L G} + \mathbf{G} \mathbf{L}^\dagger \right)_{ij} \delta (t-t')\,.
  \label{eq:fdt-onsager}
 \eeq
The theoretical framework presented in this section provides the basis of our treatment of the fluctuating DBE.

\subsection{Fluctuating Boltzmann equation}
Before analysing fluctuations in the discrete Boltzmann equation, it is worthwhile to briefly review the application of the Onsager-Machlup theory to the continuum Boltzmann equation for a dilute gas as done by Fox and Uhlenbeck \cite{fox_uhlenbeck_1970b}, since both treatments are closely connected. Fox and Uhlenbeck show that the linearized Boltzmann equation with noise can be brought into the form of Eq.(\ref{eq:lin-langevin}) above, thus enabling the use of the fluctuation-dissipation relation \eqref{eq:fdt-onsager} to obtain the variance of the noise term. These results can be immediately compared to the results of our own derivation of the noise for the case of the DBE.
While the following treatment deals with the Boltzmann equation for an ideal gas, non-ideal gas interactions can in principle be introduced by an explicit mean-field force. The contribution of the interaction force to the linearized Langevin dynamics can then be accounted for by an additional reversible contribution to the time-evolution operator of the Langevin equation \eqref{eq:lin-langevin}. However, rather than dealing with this situation for the continuum Boltzmann equation, we postpone this to section \ref{sec:FDBE}, where the discrete Boltzmann equation for a non-ideal gas is discussed. This allows us in the following to focus on the essential aspects of the treatment of the continuum Boltzmann equation.

The Boltzmann distribution function $f(\mb{r}, \mb{c}, t)$ represents the number of particles at time $t$ in a volume element $d\mb{r}d\mb{c}$ around the point $(\mb{r}, \mb{c})$ in the six-dimensional phase space of the particles. For particles of mass $\mu$ in equilibrium at temperature $T$ with a density $\rho_0$, this distribution is independent of time and the coordinate $\rv$, and has a global-Maxwellian form,
\beq
\bar f(\cv) = \rho_0 \left(\frac{\mu}{2 \pi k_B T}\right)^{d/2} \exp \left( - \frac{\mu}{2 k_B T}c^2 \right)\,,
\label{eq:fMB}
\eeq
with $d$ being the spatial dimension.
Note that, in order to remain close to the conventions used in the context of the LBE, we define the distribution function in terms of the mass density $\rho_0$ rather than the number density and take the velocity $\cv$ as the fundamental phase-space variable instead of the momentum.
Close to equilibrium, then, it is possible to expand the distribution function around its global Maxwellian form,
\beqn
f (\mathbf{r}, \mathbf{c}, t) = \bar f(\cv) \left[ 1 + h(\mathbf{r}, \cv, t)  \right]
\eeqn
and argue that $h(\mb{r}, \mb{c}, t)$ satisfies the linearized Boltzmann equation,
\beq
\partial_t h + \cv \cdot \nabla h = - \int d^d \cv' \bar f(\cv) \Lambda(\cv, \cv') \, h(\mathbf{r}, \cv', t)
\label{eq:lin_Boltzmann}
\eeq
The collision term is written using the Hilbert-Enskog kernel $\Lambda(\mb{c}, \mb{c}')$. Hilbert and Enskog have shown that the kernel is symmetric in $\mb{c}$ and $\mb{c}'$, isotropic, and has non-negative eigenvalues. Additionally, from the $d+2$ conservation laws of mass, momentum and energy, it follows that the kernel has $d+2$ null eigenvalues. The eigenspectrum of the kernel cannot, in general, be obtained analytically. However, the eigenfunctions form a complete basis for expansion of functions in velocity space. In what follows, only the knowledge of the null-space and the positive-definiteness of the remaining eigenvalues are needed. Notice that the first two terms in eq.~(\ref{eq:lin_Boltzmann}) are odd under time reversal, while the last term is even. It is this mixed character of the equation which necessitates Fox and Uhlenbeck’s generalization of the Onsager-Machlup formulation.

To transform the linear Boltzmann equation into a form consistent with Eq.(\ref{eq:lin-langevin}), define
\begin{eqnarray*}
 a(\mathbf{r}, \cv, t) &=& \sqrt{\bar f(\cv)} \, h(\mathbf{r}, \cv, t)  \\
 A (\mathbf{r}, \cv; \, \mathbf{r}', \cv') &=& \sqrt{\bar f(\cv)} \, \cv \cdot \nabla \delta (\mathbf{r} - \mathbf{r}') \, \delta (\cv - \cv') \\
 S (\mathbf{r}, \cv; \, \mathbf{r}', \cv') &=& \sqrt{\bar f(\cv) \bar f(\cv')} \Lambda(\cv, \cv') \delta (\mathbf{r} - \mathbf{r}')
\end{eqnarray*}
Treating the labels $\mb{r}$ and $\mb{c}$ as indices, $A(\mb{r}, \mb{c}; \mb{r}' , \mb{c}')$ is antisymmetric because of the factor $\nabla \delta (\mb{r} - \mb{r}')$, while $S(\mb{r}, \mb{c}; \mb{r}', \mb{c}')$ is symmetric with non-negative eigenvalues because of the factor $\Lambda(\mb{c}, \mb{c}')$. In these variables, the linearized Boltzmann equation can be written as,
\beqn
\begin{split}
 \partial_t a(\mathbf{r}, \cv, t) + \int d^d \mathbf{r}' d^d \cv' A (\mathbf{r}, \cv; \, \mathbf{r}', \cv') \, a(\mathbf{r}', \cv', t) \\
 = - \int d^d \mathbf{r}' d^d \cv' S (\mathbf{r}, \cv; \, \mathbf{r}', \cv') \, a(\mathbf{r}', \cv', t)
 \end{split}
 \label{eq:fluctCBE}
\eeqn
Noting the symmetry of the kernels, the above equation has the form of the standard regression equation of a Gaussian Markov process,
 \beqn
  \frac{d}{dt} \, \langle \mathbf{a} \rangle + \mathbf{A} \cdot \langle \mathbf{a} \rangle = - \mathbf{S} \cdot \langle \mathbf{a} \rangle
 \eeqn
where $\mb{A}$ is antisymmetric in the indices, and $\mb{S}$ is symmetric in the indices and has non-negative eigenvalues. The phenomenological theory of fluctuations in the Boltzmann equation is now reduced to adding noise terms which respect the conservation laws and reproduce Maxwell-Boltzmann equilibrium.

The linear regression equation of $a(\mb{r}, \mb{c}, t)$ is transformed into a stochastic differential equation for a Gaussian Markov process by adding a noise term $\sqrt{\bar f(\mb{c})} \, \xi(\mb{r}, \mb{c}, t)$. In the original variables, this equation gives,
\beqn
 \partial_t h + \cv \cdot \nabla h = - \int d^d \cv' \bar f(\cv) \Lambda(\cv, \cv') \, h(\mathbf{r}, \cv', t) + \xi (\mathbf{r}, \cv, t)
\eeqn
where $\xi(\mb{r}, \mb{c}, t)$ is a zero mean Gaussian random variable with a correlation that has to be determined from the FDT, eq.~\eqref{eq:fdt-onsager}. The essential step is to expand the Boltzmann entropy to quadratic order in $h(\mb{r}, \mb{c}, t)$ and obtain the entropy matrix $G(\mathbf{r},\cv; \, \mathbf{r}', \cv') \equiv \bra a(\rv,\cv) a(\rv',\cv') \ket$ which occurs in the fluctuation-dissipation relation. Fox and Uhlenbeck obtain
\beq
 G^{-1} (\mathbf{r}, \cv; \, \mathbf{r}', \cv') = \delta (\mathbf{r} - \mathbf{r}') \, \delta (\cv - \cv')\,,
 \label{eq:entropy-matrix}
\eeq
reflecting the fact that, in a dilute gas, particles are essentially uncorrelated and obey Poissonian fluctuation statistics \cite{Landau_StatPhys1}.
It follows immediately that
\beq
 \langle \xi (\mathbf{r}, \cv, t) \, \xi (\mathbf{r}', \cv', t' )  \rangle = 2 \Lambda(\cv, \cv') \, \delta (\mathbf{r} - \mathbf{r}') \, \delta(t - t')\,.
 \label{eq:cbe-noise-correlation}
\eeq
This is the fluctuation-dissipation relation for the linear fluctuating Boltzmann equation. It is intuitively clear why the variance of the fluctuation is determined by the collision kernel alone. The advection kernel contains the reversible part of the dynamics and only shifts the distribution function in phase space, without a change of shape. The collision kernel, on the other hand, causes dissipation and therefore a broadening of the distribution function. Thus, eq.~\eqref{eq:cbe-noise-correlation} means that the strength of the fluctuations is related to the amount of dissipation -- the essential content of the fluctuation-dissipation relation.

An explicit construction of the noise is now possible in terms of the eigenfunctions and eigenvalues of the collision operator. Assuming that there are discrete eigenfunctions $\phi_i(\mb{c})$ with eigenvalues $\lambda_i$, which are orthonormal with respect to the weight function $w(\mb{c}) = \bar f(\mb{c})/\rho_0$,
\beqn
 \int d \cv \, w (\cv)  \phi_i (\cv)  \phi_j (\cv) = \delta_{ij}
\eeqn
the kernel has the expansion
\beqn
 \Lambda(\cv, \cv') = \sum_{i=1}^{\infty} \lambda_i \phi_i (\cv)  \phi_i (\cv')
\eeqn
The noise can likewise be expanded in the basis of the eigenfunctions as
\beqn
 \xi (\mathbf{r}, \cv, t) = \sum_{i=1}^{\infty} \tilde{\xi}_i (\mathbf{r}, t) \phi_i (\cv)
\eeqn
The conservation laws imply the collision kernel has $d+2$ zero eigenvalues. Taking these (in three dimensions) to be $i = 1, \ldots, 5$ we see that the summations must start from $i = 6$ in the expansion of the Hilbert-Enskog kernel, and in the construction of the noise, we must set $\tilde{\xi}_i(\mb{r}, t) = 0$ for $i = 1, \ldots, 5$, since the conserved modes do not contribute to dissipation. Then, it only remains to obtain the variances of the spatial coefficients of the noise and a little more algebra shows that these are determined by the eigenvalues of the collision kernel. The final result is
\beqn
 \langle \tilde{\xi}_i (\mathbf{r}, t) \tilde{\xi}_j (\mathbf{r}', t') \rangle = 2 \, \lambda_i \, \delta_{ij} \, \delta (\mathbf{r} - \mathbf{r}') \, \delta(t - t')\,.
\eeqn
This result shows that every non-conserved eigenmode contributes to the dissipation, and therefore, every non-conserved mode must receive fluctuations from the noise to maintain Boltzmann equilibrium. The theory above provides a rigorous justification of the assumptions used intuitively in \cite{adhikari_fluct_2005}.

\subsection{Correlations in kinetic theory}
\label{sec:correl-kinetic}
As seen from the above derivation, a crucial ingredient to the FDT of the Boltzmann equation is the equilibrium correlation matrix (entropy matrix) $G=\bra h(\rv,\cv) h(\rv',\cv')\ket$ of the fluctuations $h = (f-\bar f)/\bar f$ of the distribution function.
This matrix can be determined if an expression for the equilibrium entropy in terms of the $h$'s is known, as is the case for the continuum Boltzmann equation due to the H-theorem. A more direct approach to $G$, that can moreover easily deal with non-ideal fluids, is based on the kinetic theory of fluctuations due to Klimontovich \cite{klimontovich_nonideal_1973, liboff_book, gross_pre2010}. This approach will form the basis for our treatment for fluctuations in the discrete Boltzmann equation for the non-ideal gas.

The starting point is the notion of a phase-space density $F$, defined as
\beqn
F(\rv,\cv,t)=\mu \sum_i \delta(\rv-\rv_i(t))\delta(\cv-\cv_i(t))\,,
\eeqn
where $\rv_i$ and $\cv_i$ denote the (time-dependent) positions and momenta of the individual particles of mass $\mu$ of the fluid.
As $\int F(\rv,\cv,t) d\rv d\cv/\mu=N$, the quantity $F(\rv,\cv,t) d\rv d\cv/\mu$ represents the \emph{instantaneous} number of particles in the phase-space cell $d\rv d\cv$.
The average number of particles (times mass) $\bra F \ket$ in a phase-space cell can be obtained by integrating $F$ weighted by the full $N$-particle distribution function $f^N$, $\bra F(\rv,\cv,t) \ket = \int d\rv_1 \cdots d\rv_N d\cv_1 \cdots d\cv_N f_N(\rv_1,\ldots,\rv_N,\cv_1,\ldots \cv_N) F(\rv,\cv,t)$. Obviously, this definition is equal to the one-particle distribution function $f_1$,
\beqn\bra F(\rv,\cv,t) \ket = f_1(\rv,\cv,t)\,.
\eeqn
The average of the second moment $\bra F(\rv,\cv,t) F(\rv',\cv',t')\ket$ can be related to the two-particle distribution function $f_2(\rv,\cv,\rv',\cv',t,t')$ by separating from the double-sum of delta functions the part where the two particles are identical. This gives
\beq
\bra F(\rv,\cv,t)F(\rv',\cv',t') \ket = f_2(\rv,\cv,t,\rv',\cv',t') + \mu \delta(\rv-\rv')\delta(\cv-\cv')\delta(t-t')  f_1(\rv,\cv,t)\,.
\label{eq:FF-correl}
\eeq
Similar relations can be derived for all reduced $n$-particle distribution functions. In the present work, we will only need the reduced one- and two-particle distribution functions.

The quantity $f_1(\rv,\cv,t) = \bra F(\rv,\cv,t)\ket$ denotes the ensemble-averaged number of particles (times mass) at $\rv,\cv$ at time $t$, and consequently, its time evolution will approximately be governed by the deterministic Boltzmann equation \cite{liboff_book}.
Intuitively, this notion motivates us to interpret the quantity $F$ itself as the instantaneous fluctuating one-particle distribution function $f_1^f(\rv,\cv,t)$. However, as the fluctuations in $F$ are due to collisions between a large number of particles, the assumption of molecular chaos suggests the time evolution of $f_1^f$ to be approximately governed by the Boltzmann-Langevin equation \cite{bixon_zwanzig_1969}.
In this case, the fluctuations of $f_1^f$ have Gaussian character with only their second moment being non-trivial, in contrast to the phase-space density $F$, which encapsulates also higher-order correlation functions.
The essential relation for the correlations of fluctuations of the one-particle distribution function is provided by the ansatz
\beqn \bra f_1^f(\rv,\cv,t)f_1^f(\rv',\cv',t') \ket = \bra F(\rv,\cv,t)F(\rv',\cv',t') \ket \,,\eeqn
which, as we show below, allows us to make a connection via eq.~\eqref{eq:FF-correl} to a statistical mechanical model for the equal-time pair correlation function.

For the purpose of deriving the FDT, one usually considers fluctuations around a global equilibrium state with density $\rho_0$ and zero flow velocity, i.e. $f_1$ in the previous definition is taken to be a global Maxwellian distribution, $f_1(\rv,\cv,t) = \bar f_1(\cv)$, with $\bar f$ given by eq.~\eqref{eq:fMB}. The quantity $\delta f_1(\rv,\cv,t)\equiv f_1^f(\rv,\cv,t) - \bar f_1(\cv)$ then represents the fluctuations of the one-particle distribution function $f_1$ over a uniform reference state described by $\bar f_1(\cv)$.
Restricting to a translationally invariant system, the equal-time correlation function follows as
\beq\begin{split}
\bra \delta f_1(\rv,\cv) \delta f_1(\rv',\cv')\ket &= \bra F(\rv,\cv) F(\rv',\cv')\ket -  \bar f_1(\cv) \bar f_1(\cv')\\
&= \bar f_1(\cv) \bar f_1(\cv') [g(\rv-\rv')-1]  + \mu\delta(\rv-\rv')\delta(\cv-\cv') \bar f_1(\cv)\,,
\label{eq:f1-fluct-real}
\end{split}\eeq
where, in the last step, we introduced the static pair correlation function $g$ by $f_2(\rv-\rv',\cv,\cv') = \bar f_1(\cv) \bar f_1(\cv') g(\rv-\rv')$.
As the static structure factor $S(\rv) = \bra \delta \rho(\rv_0+\rv)\delta \rho(\rv_0) \ket$ is related to the pair correlation function by \cite{Chaikin_book, Hansen_ThoSL}
\beqn S(\rv) = \rho_0^2 (g(\rv)-1) + \mu \rho_0 \delta(\rv)\,,
\eeqn
the desired correlation function for the fluctuations of the one-particle distribution function is finally obtained as
\beq
\bra \delta f(\rv,\cv) \delta f(\rv',\cv')\ket =  \Big[ \bar f(\cv) \bar f(\cv') [S(\rv-\rv')/\rho_0 - \mu \delta(\rv-\rv')]/\rho_0  + \mu \bar f(\cv) \delta(\cv-\cv')\delta(\rv-\rv') \Big]\,.
\label{eq:f-correl-cont}
\eeq
The first term on the r.h.s.\ of \eqref{eq:f-correl-cont} describes spatial correlations due to the non-ideal character of the fluid. For the ideal gas, we have $g(\rv)=1$, $S(\rv)=\mu \rho_0 \delta(\rv)$, thus only the last term remains, and relation \eqref{eq:f-correl-cont} becomes the well-known expression used in the Boltzmann-Langevin theory for a dilute gas \cite{bixon_zwanzig_1969, fox_uhlenbeck_1970b, Landau_PhysKin},
\beq \bra \delta f(\rv,\cv) \delta f(\rv',\cv')\ket =  \mu\bar f(\cv) \delta(\cv-\cv') \delta(\rv -\rv')\,.
\label{eq:f-correl-cont-ideal}
\eeq
The presence of the two delta-functions reflects the fact that in the ideal gas, there are no correlations between two particles at different positions or with different momenta. The factor $\mu$ arises due to the definition of the distribution function in terms of the mass density.

It can be checked that expression \eqref{eq:f-correl-cont} is self-consistent regarding the definition of the structure factor: As the density is the zeroth moment of the distribution function, $\rho(\rv) = \int f(\rv,\cv) d^d\cv$, we immediately obtain
$\bra \delta \rho(\rv) \delta \rho(\rv') \ket = \int d^d\cv d^d\cv' \bra \delta f(\rv,\cv) \delta f(\rv',\cv')\ket
= S(\rv-\rv')$.
Similarly, by the definition of the macroscopic fluid momentum as the first moment of the distribution function,
$ j_\alpha(\rv) = \int d^d\cv \,c_\alpha f(\rv,\cv) $, we obtain the well-known expression \cite{Landau_FluidMech59} for the equal-time momentum correlation function,
\beq \bra \delta j_\alpha(\rv) \delta j_\beta(\rv') \ket = \int d^d\cv d^d\cv'\,c_\alpha c_\beta' \bra \delta f(\rv,\cv) \delta f(\rv',\rv')\ket = \mu \int d^d\cv c_\alpha^2 \bar f(\cv) \delta(\rv-\rv')\delta_{\alpha\beta} = \rho_0 k_B T \delta(\rv-\rv')\delta_{\alpha\beta}\,,
\label{eq:momtm-correl-c}
\eeq
Here, we used the fact that $\bar f$ [eq.~\eqref{eq:fMB}] describes a quiescent state, i.e.\ $\int d^d\cv\,\cv  \bar f(\cv) =\bv{0}$, and the Gaussian integral $\int d^d\bv{c} \bar f(\bv{c}) c_\alpha c_\beta =  \delta_{\alpha \beta}\rho_0 k_B T / \mu$.
Expression \eqref{eq:momtm-correl-c} shows that, both in the ideal and non-ideal fluid, there are no correlations in the momenta of different fluid elements. Indeed, as is well-known from statistical mechanics, the Hamiltonian of an equilibrium fluid is diagonal in the momenta.

\section{Discrete velocity Boltzmann equation}

In the derivation of the lattice Boltzmann equation, the kinetic space is discretized in the velocities to yield a discrete velocity Boltzmann equation for the distribution function $f(\mb{r}, \mb{c}, t)$ \cite{benzi_physrep1992}.
Choosing an appropriate set of discrete velocities $\mb{c}_i$, the discrete velocity Boltzmann equation for the distribution function $f_i(\mb{r}, t) \equiv f(\mb{r}, \mb{c}_i, t)$ under the action of an arbitrary body force $\Fv$ reads
\beq
 \partial_t f_i + \mathbf{c}_i \cdot \nabla f_i + [\Fv\cdot \nabla_\cv f]_i= - \Lambda_{ij} ( f_j - f_j\ueq )\,.
 \label{eq:dbe}
\eeq
Here, $f_i(\mb{r}, t)$ represents the mean number of particles at $\mb{r}$ and time $t$, moving along the lattice direction defined by the discrete velocity $\mb{c}_i$, ($i, j = 1, \ldots ,N$). Furthermore,
\beq
f_i\ueq = w_i \left( \rho + \frac{\rho \mathbf{v} \cdot \mathbf{c}_i}{\cslb^2} + \frac{\rho \, v_{\alpha} v_{\beta} \, Q_{i \alpha \beta}}{2 \, \cslb^4}  \right)
\label{eq:feq}
\eeq
is a local equilibrium distribution, the discrete analogue of a local Maxwellian distribution in continuum kinetic theory truncated to second order in the mean flow velocity $\mb{v}$, the $w_i$ are a set of weights which satisfy $\sum_i w_i = 1$, and $Q_{i\alpha \beta} = c_{i \alpha} c_{i \beta} - \cslb^2 \delta_{\alpha \beta}$.
Greek indices denote Cartesian directions and summation over repeated free indices is implied.
The constant $\cslb$ is the speed of sound of an \emph{ideal} LB gas, and its numerical value is fixed by the structure of the discrete velocity set. For the conventional D2Q9 lattice employed below, it has a numerical value of $\cslb^2=1/3$ in lattice units. Note that this quantity has to be distinguished from the actual speed of sound $c_s$ of the fluid under consideration. If the fluid is non-ideal, $c_s$ will in general be different from $\cslb$.
The quantity $[\Fv\cdot \nabla_\cv f]_i$ represents the discretized equivalent of a forcing term, which we take as \cite{martys_force_1998}
\beq \Phi_i \equiv -[\Fv\cdot \nabla_\cv f]_i = \rho w_i\left(\frac{\Fv\cdot \cv_i}{\cslb^2} + \frac{(\vv \Fv + \Fv\vv): \bv{Q}_i}{2\cslb^2}\right)\,.
\label{eq:forcing-term}
\eeq
For present purposes, where a body force is supposed to mediate the non-ideal fluid interactions in the fluctuating DBE, we will assume that $\Fv$ depends only on the density and its derivatives, $\Fv=\Fv(\rho, \nabla\rho)$.

The low order velocity moments of the distribution function are related to the densities of mass, momentum and the deviatoric stress,
\beq\{\rho, \rho v_{\alpha}, S_{\alpha \beta} \} =
\sum_i f_i \{ 1, c_{i\alpha}, Q_{i\alpha \beta} \}
\label{eq:f-moments}
\eeq
where $S_{\alpha \beta} + \rho \cslb^2 \delta_{\alpha \beta} = \Pi_{\alpha \beta}$ is the
Eulerian momentum flux. The higher moments of the distribution function are related to the densities of rapidly relaxing kinetic degrees of freedom, variously called ghost or kinetic variables. In eq.~\eqref{eq:dbe}, $\Lambda_{ij}$ is a scattering matrix whose eigenvalues control the relaxation of the kinetic modes to their local equilibrium values \cite{higuera_epl1989, higuera_succi_epl1989}. The null eigenvalues correspond to the eigenvectors associated with the conserved mass and momentum densities, while the leading non-zero eigenvalue associated with $Q_{i\alpha \beta}$ controls the viscosity of the LB fluid.

For a general athermal DBE model with $n$ velocities in $d$ space dimensions, the $n \times n$ collision matrix $\Lambda_{ij}$ has $d+1$ null eigenvectors corresponding to the conserved density and $d$ components of the conserved momentum, $d(d + 1)/2$ eigenvectors corresponding to the stress modes, and $n - (d + 1) - d(d + 1)/2$ eigenvectors corresponding to the ghost modes \cite{dHumieres_MRT_1992, benzi_physrep1992}. The choice of the null and stress eigenvectors $\{ 1, c_{i\alpha}, Q_{i\alpha \beta} \}$ follows directly from the physical definition of the densities associated with them. Without specifying the exact analytical expression for the remaining eigenvectors, let us assume a linearly independent set of the eigenvectors of the scattering matrix be given by $\{T_{ai}\}$, where $a = 1 \ldots n$ labels the eigenvector, and $i = 1 \ldots n$ labels the component of the eigenvector along the $i$−th velocity direction.
This allows us to define densities associated with the eigenvectors $T_{a}$ as moments of the populations by
\beq
 m_a (\rv, t) =  T_{ai} f_i(\rv, t)  \,.
 \label{eq:moments}
\eeq
For $T_{ai}\in \{ 1, c_{i\alpha}, Q_{i\alpha \beta} \}$, the densities are the mass, momentum and stress. The ghost eigenvectors are higher polynomials of the discrete velocities. The discreteness of the kinetic space implies that, unlike in the continuum, only a finite number of polynomials can be linearly independent, being equal to the number of discrete velocities. For a model with $n$ discrete velocities, the choice of the $n$ linearly independent polynomials is not unique, but defined only up to a similarity transformation.
Independent of the precise choice, the distribution function itself can be expanded in a linearly independent set of eigenvectors which are polynomials of the discrete velocities
\beq
 f_i(\rv, t) = w_i  \frac{T_{ai}}{N_a} m_a(\rv, t) \,.
 \label{eq:moments-inv}
\eeq
Consistency between the above two equations implies that the set of polynomials $T_{a}$ are both orthogonal and complete,
\begin{eqnarray}
 \sum_i w_i T_{ai} T_{bi} &=& N_a \delta_{ab} \label{eq:T-orthog}\,,\\
 w_i\sum_a \frac{T_{ai} T_{aj}}    {N_a} &=& \delta_{ij} \,,
 \label{eq:T-complete}
\end{eqnarray}
where $N_a$ is the length of the $a$th eigenvector.
Crucially, with the definitions above, the eigenvectors $T_{ai}$ form an orthogonal set under a weighted inner product $(T_a,T_b) = \sum_i w_i T_{ai} T_{bi}$ \cite{adhikari_fluct_2005}.
Compared to using an unweighted inner product \cite{dHumieres_MRT_1992, dHumieres_MRT_2002}, the advantage of the present choice is that the equilibrium distribution \eqref{eq:feq} has no projection onto the ghost sector, i.e.\ $T_{ai} f_i\ueq=0$ if $T_a$ is a ghost eigenvector. This emphasizes the physical interpretation of the model, but is otherwise not necessary for a faithful implementation of hydrodynamics \cite{kaehler_mrt_2010}.
It is often convenient to imagine the $f_i$ to be the elements of vector in the space spanned by the velocity vectors $\cv_i$. Then, $T_{ai}$ defines via \eqref{eq:moments} an orthogonal transformation matrix between the distribution function-space and the moment-space. Relation \eqref{eq:moments-inv} serves to define the transformation matrix in the inverse direction, $(T^{-1})_{ia}=w_i T_{ai}/N_a$. One possible choice for the $T_a$ for a D2Q9 lattice is presented in appendix \ref{sec:basis-set}.

Based on the notion that $T_a$ represents polynomials corresponding to the hydrodynamic, transport, and ghost eigenvectors, correspondingly, eq.~\eqref{eq:moments-inv} allows the distribution function to be separated into contributions from the hydrodynamic, transport, and ghost moments
\beq
 f_i = f_i^H + f_i^T + f_i^G\,.
 \label{eq:split_dist_fn}
\eeq
This structure can be conveniently organized by introducing projection operators \cite{behrend_hydro_1994} which project the distribution function onto the hydrodynamic, transport, and ghost subspaces,
\begin{eqnarray}
 P_{ij}^H f_j &=& f_i^H = w_i \left( \rho + \frac{\rho \mathbf{v} \cdot \mathbf{c}_i}{\cslb^2} \right)\,,  \\
 P_{ij}^T f_j &=& f_i^T = w_i \, \frac{S_{\alpha \beta} Q_{i \alpha \beta}}{2 \cslb^4}\,,    \\
 P_{ij}^G f_j &=& f_i^G = w_i \sum_{a \in G} m_a \frac{T_{ai}}{N^a}\,.
 \label{eq:proj_operators}
\end{eqnarray}
The explicit form of the projection operators immediately follow from the completeness relation \eqref{eq:T-complete} as
\beqn
 P_{ij}^H = w_i \left( 1 + \frac{\mathbf{c}_i \cdot \mathbf{c_j} }{\cslb^2}  \right)\,,\qquad
 P_{ij}^T = \frac{w_i  Q_{i \alpha \beta}  Q_{j \alpha \beta}}{2\, \cslb^4} \,,\qquad
 P_{ij}^G = \sum_{a \in G} \frac{w_i T_{ai} T_{aj}}{N^a}\,.
\eeqn

The discrete Maxwellian and the forcing term is a nonlinear function of the distribution function, and thus eq.~(\ref{eq:dbe}) is only apparently linear, the nonlinearity being concealed in $f_i\ueq$ and $\Phi_i$. A useful linearization of the LB equation consists of neglecting the quadratic term in the discrete Maxwellian to yield a local equilibrium $h_i\ueq$ which is linear in the mean velocity,
\beq
h_i\ueq \equiv w_i\rho \left( 1 + \frac{\mathbf{v} \cdot \mathbf{c}_i}{ \cslb^2} \right) = P_{ij}^H f_j\,.
\label{eq:lin-feq}
\eeq
Similarly, the forcing term $\Phi_i$, eq.~\eqref{eq:forcing-term}, can be linearized as
\beq
\phi_i \equiv  P_{ij}^H \Phi_j(\delta \Fv) = \rho w_i \frac{\delta \Fv\cdot \cv_i}{\cslb^2} \equiv K_{ij} f_j\,.
\label{eq:lin-force}
\eeq
where the last equality serves to define the linear forcing operator $K_{ij}$. This definition is always possible, as the linearized interaction force $\delta \Fv$ is assumed to depend linearly on the density and its derivatives.
While the definitions of $\delta \Fv$ and $K_{ij}$ are formal at the level of eq.~\eqref{eq:lin-force}, their precise forms will become clear in the following section, where we discuss a particular DBE model for a non-ideal fluid.
In the linearized approximation for the equilibrium distribution, we have $f_i\ueq = h_i\ueq = (P^Hf)_i$.
Defining a right-projected relaxation matrix as $\Lambda^R_{ij} \equiv \Lambda_{ik}(1-P^H)_{kj}$, the relaxation term of the DBE \eqref{eq:dbe} can be written as
\beqn \Lambda_{ij}\left[f_j - f\ueq_j \right] = \Lambda_{ij}\left[f_j - (P^H f)_j \right] = \Lambda_{ij}^R f_j\,,
\eeqn
which is obviously linear in $f_i$.
Thus, the linearized DBE follows as
\beq
 \partial_t \delta f_i + \mathbf{c}_i \cdot \nabla \delta f_i = - \Lambda_{ij}^R \delta f_j + K_{ij} \delta f_j
 \label{eq:lin-DBE}
\eeq
where the obvious notation $\delta f_i$ is introduced to indicate a fluctuation.

It is clear that $\Lambda^R_{ij}$ by construction has eigenvectors of mass and momentum with zero eigenvalues. The form of the matrix, by itself, places no constraint on the eigenvalues of the transport and ghost sectors.
The simplest possible choice consists of the BGK approximation, where the relaxation matrix is diagonal, $\Lambda_{ij} = \delta_{ij}/\tau$, which implies that all non-conserved modes relax at the same rate $1/\tau$.
In general, the maximum number of different eigenvalues depends on the dimensionality and the chosen basis set $T_a$ \cite{succi_book}. The fluctuating DBE that we will present below places no further constraints on the eigenvalues, and we will denote the eigenvalues of $\Lambda_{ij}^R$ simply by $\lambda_a$, i.e.\
\beq \Lambda_{ij}^R T_{aj} = \lambda_a T_{ai}\,, \label{eq:relax-eigenv} \eeq
keeping in mind that by construction $\lambda_a=0$ for the conserved modes (i.e.\ $a=1,\ldots,d+1$).

The linearized DBE \eqref{eq:lin-DBE} is most conveniently analysed in Fourier space. Introducing the Fourier transformation by $\delta f_i(\rv) = (1/2\pi)^{d/2} \int \delta f_i(\kv) \exp(-\im \kv\cdot \rv)\,d\kv $ yields
\beq
 \partial_t \delta f_i(\kv) - \im \mathbf{k} \cdot \mathbf{c}_i \delta f_i(\kv) = \left[- \Lambda_{ij}^R  + K_{ij}(\kv)\right]\delta f_j(\kv) \,.
 \label{eq:lin-dbe-fourier}
\eeq
Thus, the DBE dynamics is reduced to a set of coupled ordinary differential equations for the Fourier modes $\delta f_i(\mb{k},t)$.
The linearized DBE can equally well be written in terms of the Fourier transforms of the moments $\delta m_a(\mb{k}, t)$, using definition \eqref{eq:moments} to obtain
\beq
 \partial_t \delta m_a(\kv) + A_{ab}(\mathbf{k}) \delta m_b(\kv)=  \left[ - \lambda_a \delta_{ab} + \hat K_{ab}(\kv) \right] \delta m_b(\kv)\,,
 \label{eq:lin-dbe-mom}
\eeq
where $A_{ab}(\kv)$ is the Fourier-transformed advection operator (see appendix \ref{sec:adv})
\beq
 A_{ab}(\kv) = -\im \kv\cdot  T_{ai} \cv_i T^{-1}_{ib} = -\im \mathbf{k} \cdot w_i T_{ai} T_{bi} \mathbf{c}_i/N_a\,,
 \label{eq:adv-fourier}
\eeq
and $\hat K_{ab}=T_{ai}K_{ij} T^{-1}_{ib}$ denotes the forcing operator in moment space.
Clearly, since the populations and densities are related by an invertible linear transformation, eq.~\eqref{eq:lin-dbe-mom} represents an exact reformulation of the DBE dynamics in the basis of moments.
The dynamical equation in the $f_i$ basis diagonalizes the advection operator $ -\im\mb{k} \cdot \mb{c}_i\delta_{ij}$, while the dynamical equation in the $m_a$ basis diagonalizes the collision operator $\Lambda^R_{ij}$ . The eigenvectors of the dynamics are a combination of the $f_i$ and the $m_a$. A forcing term will contribute off-diagonal terms in either representation, as the $m_a$ are constructed as eigenvectors of the relaxational part of the collision operator, $\Lambda^R_{ij}$.

\section{Fluctuating Discrete-Velocity Boltzmann Equation}
\label{sec:FDBE}
We shall now turn to the central result of the present work: the fluctuating DBE and the corresponding fluctuation-dissipation relation.
The fluctuating DBE is obtained by promoting the DBE \eqref{eq:lin-DBE} into a Langevin equation, where the $f_i$ are interpreted as instantaneous, fluctuating densities in phase space (corresponding to the $f_1^f$ in the notation of section \ref{sec:correl-kinetic}),
\beqn
 \partial_t f_i + \mathbf{c}_i \cdot \nabla f_i = - \Lambda_{ij} ( f_j - f_j\ueq ) + \Phi_i + \xi_i\,.
\eeqn
The noise terms $\xi_i$ are Gaussian random variables that give rise to fluctuations in the populations in each phase space cell.
Restricting to linear fluctuations, the FDBE becomes
\beq
 \partial_t \delta f_i + \mathbf{c}_i \cdot \nabla \delta f_i = - \Lambda_{ij}^R \delta f_j + K_{ij}\delta f_j + \xi_i \,,
 \label{eq:FDBE-f}
\eeq
where, by construction, the linear relaxation matrix $\Lambda^R_{ij}$ has zero eigenvalues for the conserved modes [see eq.~\eqref{eq:relax-eigenv}].
After Fourier-transforming the linear FDBE \eqref{eq:FDBE-f} and expressing it in terms of moments, we obtain
\beq\begin{split}
 \partial_t \delta m_a(\kv) &=  \left[- A_{ab}(\mathbf{k}) - \lambda_a \delta_{ab} + \hat K_{ab}(\kv) \right] \delta m_b(\kv) + \hat \xi_a(\kv) \\
&= -L_{ab}(\kv) \delta m_b(\kv) + \hat \xi_a(\kv)\,,
 \label{eq:FDBE-m}
\end{split}\eeq
with $\hat \xi_a=T_{ai} \xi_i$. For later convenience, we introduced the DBE time-evolution operator $\mathbf{L}$ by
\beqn L_{ab}(\kv) \equiv A_{ab}(\mathbf{k}) + \lambda_a \delta_{ab} - \hat K_{ab}(\kv)\,.\eeqn
Above linear FDBE can be brought into the form of Langevin eq.~\eqref{eq:lin-langevin} by writing it as
\beqn \partial_t \delta m_a(\kv) = -\int d \kv' \mathcal{L}_{ab}(\kv,\kv') m_b(\kv') + \hat \xi_a(\kv)\,,
\eeqn
with $\mathcal{L}_{ab}(\kv,\kv')\equiv L_{ab}(\kv)\delta(\kv-\kv')$, and interpreting the wavenumber arguments formally as indices. Hence, the operator defined by $\lambda_a\delta_{ab}\delta(\kv-\kv')$ is Hermitian in the wavenumber and mode indices, while the operator defined by $A_{ab}(\kv)\delta(\kv-\kv')$ is anti-Hermitian in the wavenumber indices [cf.~\eqref{eq:adv-fourier}], but of mixed character in the mode indices.
As will be shown below, the same holds for the operator $\hat K_{ab}(\kv)\delta(\kv-\kv')$.
Denoting the equal-time correlation matrix of the modes by $\mathcal{G}_{ab}(\kv,\kv') = \bra m_a(\kv) m_b(\kv')\ket $,
the noise correlations follow straightforwardly from the FDT, eq.~\eqref{eq:fdt-onsager}, as
\beq
\bra \hat \xi_a (\kv, t) \, \hat \xi_b (\kv', t' )  \ket = \int d\kv'' \left[\mathcal{L}(\kv,\kv'') \mathcal{G}(\kv'',\kv') + \mathcal{G}(\kv,\kv'') \mathcal{L}^\dagger(\kv'',\kv') \right]_{ab} \delta(t - t') \,.
\label{eq:dbe-fdt-gen}
\eeq
The above relation can be simplified by noting that, due to translational invariance, the mode correlations are generally diagonal in Fourier space, i.e.\ $\mathcal{G}(\kv,\kv') \equiv G(\kv)\delta(\kv+\kv')$ with $G_{ab}(\kv) \equiv \bra m_a(\kv) m_b(-\kv)\ket = \bra m_a(\kv) m^*_b(\kv)\ket $.
This allows to obtain the FDT in its final form
\beq
\Xi_{ab} (\kv) \equiv \bra \hat \xi_a (\kv) \, \hat \xi_b (-\kv) \ket  = \left[\mathbf{L}(\kv) \mathbf{G}(\kv) + \mathbf{G}(\kv) \mathbf{L}^\dagger(\kv) \right]_{ab} \,,
\label{eq:dbe-fdt}
\eeq
where we defined the equal-time correlation matrix of the noise $\Xi_{ab}(\kv)$.
Within our convention, $G_{11}$ is the equal-time density correlation function (static structure factor) and $G_{aa}$ for $a=2,\ldots,1+d$ is the equal-time momentum correlation function.

In contrast to the time-evolution operator $\mathbf{L}$, which is provided by the DBE, the correlation matrix $\mathbf{G}$ must in general instead be derived from a statistical mechanical framework.
The specification of $\bG$ constitutes a central ingredient of any FDBE model.
While the matrix elements of the conserved subspace of $\bG$ can immediately be determined from the basic statistical mechanical theory of fluids \cite{Landau_FluidMech59}, the specification of the remaining elements is less straightforward.
In particular, it must be emphasized that the correlations of the non-conserved modes are not arbitrary, as the full matrix $\bG$ must combine in the FDT \eqref{eq:dbe-fdt} with the reversible and irreversible parts of the DBE dynamics (as represented by $\bL$) to obtain a positive-semidefinite noise covariance matrix for all wavenumbers. The intimate relation between equilibrium correlations and dynamical equations can be seen from the following argument based on the FDT of the fluctuating Navier-Stokes equations of a non-ideal fluid \cite{hohenberg_halperin, mazenko_book, gross_pre2010}: due to the coupling between density and momentum fluctuations, once the expression for the random stress tensor correlations has been found, the structure factor can be determined from the dynamical equations without any further reference to statistical mechanics.
This illustrates the general fact that driving forces in the dynamical equations are provided by thermodynamic forces and therefore, already contain information on the equilibrium fluctuations \cite{Landau_StatPhys1, deGroot_Mazur}. Transferred to the DBE, this implies that the expression for the structure factor is related to the interaction force and, more generally, the equilibrium correlations of the conserved modes are indirectly fixed by the correlations of the non-conserved modes due to the dynamic coupling between both.
As shown in \cite{duenweg_statmechLB_2007} in the context of the LBE, in principle, it suffices to derive a fluctuation model for the non-conserved modes and obtain an expression for the noise based only on this information using a detailed balance argument. In the context of the Langevin approach to the DBE, however, it will be necessary to specify the full equal-time correlation matrix of all modes, paying special attention to the relation between equilibrium correlations and interaction forces.

Finally, it should be noted that the form of the interaction force is in principle restricted by the numerics of the streaming process. This can be seen, for instance, by considering a situation of coexisting phases in a quiescent equilibrium fluid: in order to maintain the Maxwellian equilibrium distribution (with spatially varying density) at each point in space, the non-zero contribution of the forcing term in the collision step of the Boltzmann equation must be compensated solely by the streaming term, as the collision term identically vanishes. This requirement becomes delicate to fulfill in the LBE where space is discretized and demands a careful implementation of the forcing term.

In this work, we adopt eq.~\eqref{eq:f-correl-cont}, which is directly based on continuum kinetic theory, as the defining relation for the equal-time correlations of the distribution function in both the ideal and non-ideal fluid.
Written in terms of the discrete velocities, eq.~\eqref{eq:f-correl-cont} becomes in Fourier-space
\beq
\bra \delta f_i(\kv) \delta f_j(\kv')\ket =  \Big[ \bar f_i \bar f_j [S(\kv)/\rho_0-\mu]/\rho_0  + \mu\bar f_i \delta_{ij} \Big]\delta(\kv+\kv')\,,
\label{eq:f-correl-dbe}
\eeq
where $\bar f_i(\cv) = f_i\ueq(\cv,\rho_0, \vv=0)$ is the discrete Maxwellian. 
Due to translational invariance, all correlation functions are diagonal in Fourier-space, thus allowing us to write them in terms of a single wavevector.
The moment-space correlation matrix is obtained by projecting \eqref{eq:f-correl-dbe} onto the basis vectors,
\beq \begin{split}
G_{ab}(\kv) &= \bra \delta m_a(\kv) \delta m_b(-\kv)\ket = T_{ai} T_{bj} \bra \delta f_i(\kv) \delta f_j(-\kv) \ket \\
&= \left(\bar m_a \bar m_b [S(\kv)/\rho_0-\mu]/\rho_0 + \mu T_{ai} T_{bi} \bar f_i\right)\,,
\label{eq:g-correl}
\end{split}\eeq
where $\bar m_a = T_{ai} \bar f_i$. 
Whenever the equilibrium distribution is of ideal gas form, eq.~\eqref{eq:feq}, we have $\bar m_a = \rho_0 \delta_{a1}$.
In the above relations we have used the fact that for the Fourier-transform of a real-valued quantity, such as the distribution function or its moment, complex conjugation is equivalent to changing the sign of the wavevector.

The structure factor $S(\kv)$ in general depends on the underlying (non-ideal fluid) interactions and must itself be determined from a statistical mechanical theory, i.e.\ it is not provided by eq.~\eqref{eq:f-correl-dbe}. Note however, that for the force-based DBE to be discussed below (section \ref{sec:force-nonideal}), the structure factor will drop out of the final expression for the noise covariance $\Xi$ due to the reversible nature of the interaction force. In contrast, this will not be the case for the modified-equilibrium model (see section \ref{sec:yeo-dbe}).

Crucially, in contrast to the continuum Boltzmann case, the parameter $\mu$ representing the mass of the fluid particles is not known anymore \emph{a priori} in the algorithm, but now must instead be determined self-consistently from the expression for the momentum correlation function obtained from eq.~\eqref{eq:f-correl-dbe}. 
The discrete momentum correlation function of any fluid in equilibrium at a temperature $T$ and density $\rho_0$ is assumed to be given by the same relation \eqref{eq:momtm-correl-c} as in the continuum,
\beq \bra \delta j_\alpha(\kv) \delta j_\beta(\kv') \ket = \rho_0 k_B T  \delta_{\alpha\beta}\delta(\kv+\kv')\,,
\label{eq:momtm-correl-l}
\eeq
where the momentum density is defined as $\bv{j} = f_i \cv_i$ [see eq.~\eqref{eq:f-moments}].
From this relation, we will find that $\mu$ is related to the fluctuation temperature of the fluid, in agreement with a previous analysis of the fluctuating ideal gas LBE \cite{duenweg_statmechLB_2007}.
This is explicitly shown below (secs.~\ref{sec:fdbe-ideal} and \ref{sec:fdbe-nonideal}), where we discuss the ideal and non-ideal fluid models. 
Thus, once the structure factor and the equilibrium distribution are specified, relation \eqref{eq:g-correl} allows to compute any correlation function of two modes.

In the FDT, eq.~\eqref{eq:dbe-fdt}, it is assumed that the noise is not correlated in time, i.e.\ the $\xi_a$ represent Gaussian white noise in time. This is physically well justified, as the noise represents the ``fast''degrees of freedom of the many-particle system which typically have correlation times many orders of magnitude smaller than the interesting macroscopic quantities \cite{zwanzig_book, tremblay_noneq_1981}. Similarly, one might reason that the noise should also not be correlated in space, as the fast variables relax on scales much smaller than inherent to the description of the Boltzmann equation. Spatially uncorrelated noise will indeed be an exact result of the FDT for the force-based models discussed below.
The model of Swift et al. \cite{swift_lattice_1995}, however, deviates from this basic intuition in that it requires spatially correlated noise, as shown in \cite{gross_pre2010}. The origin of this result can be traced back to the use of a modified equilibrium distribution, which is also responsible for the non-local (wavelength-dependent) bulk viscosity arising in the associated Navier-Stokes equations of that model.
Besides the absence of temporal correlations, no further constraints will be imposed on the noise. In particular, the laws of mass and momentum conservation are not enforced a priori, but will instead be an outcome of the FDT by construction of the relaxation matrix $\Lambda^R$, as shown below.

Finally, we shall make some comments on the applicability of the noise derived from a linear theory to situations where \emph{non-linear} terms are present. As can be shown on general grounds via a Fokker-Planck treatment \cite{KimMazenko_1991, mazenko_book, Chaikin_book}, in cases where a well-defined equilibrium distribution (e.g. in terms of a free-energy functional) exists, both the noise obtained from the linear and fully non-linear hydrodynamical treatment are identical.
Thus, the noise expressions derived for the models considered in this work are expected to remain valid in the presence of non-linear couplings, provided they are either fully reversible or stem from an underlying free-energy functional. The advective non-linearity $j_\alpha j_\beta/\rho$ of hydrodynamics belongs to the first class, while cubic (or higher order) density terms in the interaction force (equation of state) represent so-called dissipative couplings originating from a free-energy functional \cite{hohenberg_halperin}. Although a strict treatment should deal with the appropriate Fokker-Planck equation for the FDBE, we nevertheless expect that the noise derived in the present work can be used to model fluctuating non-linear fluids, as, for example, required for phase coexistence or critical phenomena.

Under general \emph{non-equilibrium} conditions, as, for example, a fluid under flow, the situation is not so straightforward.
Here, it has been shown that the equations of fluctuating hydrodynamics remain valid in simple non-equilibrium steady states (e.g., in a temperature gradient or uniform shear), provided that the noise covariances are computed with the local values for the thermodynamic variables and transport coefficients \cite{keizer_fluct_1978, ronis_flhyd_1980, tremblay_noneq_1981, kirkpatrick_basic_1982, tremblay_review_1984, garcia_shear_1987, schmitz_physrep_1988, sengers_book}.
This is intuitively clear from the fact that the fast variables are uncorrelated on macroscopic length and time scales and can thus be expected to be in a local equilibrium.
Thus, it seems reasonable to assume that FDBE noise with a covariance of the same form as in the equilibrium case can also be employed in simple non-equilibrium states. However, we are not aware of any studies investigating these issues in detail for the FDBE or FLBE at the present time. We remark that violation of Galilean invariance appears to limit the application of the FLBE to systems with rather small mean flow velocities \cite{kaehler_gal_2010}.
In this work, we will be only concerned with fluctuations around equilibrium states.

\subsection{Ideal Gas}
\label{sec:fdbe-ideal}
As a first application of the theory presented above, we derive the noise covariance for the ideal gas model. Our result, obtained for the FDBE, will be shown in section \ref{sec:flbe} to be identical to the result derived by \cite{adhikari_fluct_2005} using the direct FLBE approach.
The FDBE of the ideal gas is given by
\beq
\partial_t \delta m_a(\kv) + A_{ab}(\mathbf{k}) \delta m_b(\kv) = - \lambda_a \delta m_a(\kv) + \hat \xi_a(\kv) \,,
\eeq
and, thus, the generalized time-evolution operator of eq.~\eqref{eq:FDBE-m} consists in this case only of an advective and a relaxational part,
\beq L_{ab}(\kv)  = A_{ab}(\mathbf{k}) + \lambda_a \delta_{ab}\,,\label{eq:L-ideal}\eeq
where $\lambda_{a=1,\ldots,d+1}=0$ by construction of the relaxation operator.
To compute the equilibrium correlation matrix of the modes $\bG$, we assume the structure factor to be given by the conventional (wavelength-independent) ideal gas expression $S\st{id}(\kv) = \rho_0 \mu$, with a yet undetermined mass parameter $\mu$. Hence, we obtain from eq.~\eqref{eq:f-correl-dbe} the DBE analog of relation \eqref{eq:f-correl-cont-ideal} as
\beq
\bra \delta f_i(\kv) \delta f_j(\kv') \ket = \mu \bar f_i \delta_{ij} \delta(\kv+\kv')\,,
\label{eq:f-correl-ideal}
\eeq
where $\bar f_i = \rho_0 w_i$ is the global Maxwellian.
Clearly, the independence of \eqref{eq:f-correl-ideal} of the wavenumber expresses the absence of all correlations in an ideal gas.
With the help of the orthogonality relation \eqref{eq:T-orthog}, we compute the equilibrium correlation matrix as
\beq
G_{ab}(\kv) = \bra\delta m_a(\kv) \delta m_b(-\kv) \ket = T_{ai} T_{bj}\bra \delta f_i(\kv) \delta f_j(-\kv) \ket = \mu \rho_0 w_i  T_{ai} T_{bi} = \mu \rho_0 N_a \delta_{ab} \,,
\label{eq:G-ideal}
\eeq
where $N_a$ is the length of the $a$th basis vector.
The parameter $\mu$ can now be determined by requiring consistency with the basic statistical mechanical relation \eqref{eq:momtm-correl-l} for the momentum correlation, fixing
\beq \mu = \frac{k_B T}{\cslb^2}\,,
\label{eq:mu-ideal}
\eeq
since $N_{2,\ldots,d+1} = \cslb^2$ (c.f.\ Table~\ref{tab:modes-d2q9}). As $\cslb$ is also the speed of sound of the ideal gas, we see that \eqref{eq:mu-ideal} is consistent with the ideal gas equation of state if $\mu$ is interpreted as the mass of a fictitious DBE-particle \cite{duenweg_statmechLB_2007}. From \eqref{eq:G-ideal} together with the above value of $\mu$ the complete expression for the structure factor of the ideal gas is obtained as
\beq S\st{id}(\kv) = \frac{\rho_0 k_B T}{\cslb^2}\,.\label{eq:S-ideal}
\eeq
With relations \eqref{eq:L-ideal} and \eqref{eq:G-ideal}, the noise covariance matrix finally follows from eq.~\eqref{eq:dbe-fdt} as
\beq \Xi_{ab}(\kv) = 2\lambda_a G_{ab}(\kv) = 2\lambda_a \frac{\rho_0 k_B T}{\cslb^2} N_a \delta_{ab}\,.
\label{eq:noise-ideal}
\eeq
Transforming back into real-space and invoking the definition of $\Xi_{ab}(\kv)$ as $\bra \xi_a(\kv)\xi_b(\kv')\ket = \Xi_{ab}(\kv)\delta(\kv+\kv')$, we obtain
\beqn \bra \xi_a (\rv) \xi_b(\rv')\ket = 2\lambda_a \frac{\rho_0 k_B T}{\cslb^2} N_a \delta_{ab} \delta(\rv-\rv')\,,
\eeqn
which explicitly indicates the absence of spatial correlations.
Crucially, as $\lambda_a=0$ by construction for the conserved modes [see eq.~\eqref{eq:relax-eigenv}], noise with the above covariance automatically obeys mass and momentum conservation, i.e.\ $\xi_a=0$ for $a=1,\ldots,d+1$.
Notably, the equilibrium correlation matrix of the ideal gas and the advection operator fulfil $\bA \bG = -\bG \bA^\dagger$, which is the technical reason why the above noise covariance is independent of any advective contribution, in line with physical expectation. This was also found in \cite{adhikari_fluct_2005} using the FLBE approach to the ideal gas.

\subsection{Non-Ideal Gas}
\label{sec:fdbe-nonideal}
We now proceed to analyse thermal noise in the DBE of two non-ideal gas models. First, as a central new result, a general force-based FDBE is presented that applies, among others, to the model introduced by He, Shan and Doolen \cite{he_dbe_1998}.
Next, the modified-equilibrium model of Swift et al.~\cite{swift_lattice_1995, swift_lattice_1996} is analysed within the DBE approach, thus complementing previous work based on the LBE \cite{gross_pre2010}.
The equations of fluctuating hydrodynamics of non-ideal fluids resulting from the two FDBEs below can be found in \cite{gross_pre2010}.
\subsubsection{Force-based model}
\label{sec:force-nonideal}
The general FDBE for a force-based non-ideal fluid model is given as
\beq
 \partial_t \delta m_a(\kv) + A_{ab}(\mathbf{k}) \delta m_b(\kv) =  \left[ - \lambda_a \delta_{ab} + \hat K_{ab}(\kv) \right] \delta m_b(\kv) + \hat \xi_a(\kv)\,,
\label{eq:fdbe-force}
\eeq
where $\hat K_{ab}$ is the linearized forcing-term defined through $\hat K_{ab} m_b = T_{ai} \phi_i$ with $\phi_i$ given by eq.~\eqref{eq:lin-force}, $\phi_i=\rho_0 w_i \delta \Fv\uint\cdot \cv/\cslb^2$.
The equilibrium distribution is assumed to have the standard Maxwellian form, eq.~\eqref{eq:feq}.
In the linear approximation, the forcing-term only affects the momentum density, since
\beq T_{ai}\phi_i = \{0, \delta F_\alpha,0,\ldots\}\,.
\label{eq:force-mom}
\eeq
To proceed, an expression for the interaction force in terms of the density has to be specified.
We shall assume the linear interaction force to be given by
\beq \delta \Fv\ut{int} = \im \kv \left[c_s^2(\kv) -\cslb^2\right]\delta \rho\,,
\label{eq:lin-intforce}
\eeq
where $c_s(\kv)$ is a generalized speed of sound that is related to the structure factor of the non-ideal fluid by
\beq S(\kv) = \frac{\rho_0 k_B T}{c_s^2(\kv)}\,.
\label{eq:s-nonideal}
\eeq
It is shown in appendix \ref{sec:force-models} that assumptions \eqref{eq:lin-intforce} and \eqref{eq:s-nonideal} hold, in particular, for the model of He, Shan and Doolen \cite{he_dbe_1998}.
Crucially, for the purpose of deriving the FDT, the specific expressions of $c_s(\kv)$ or $S(\kv)$ are not important.
Hence, the present model is not restricted to a particular thermodynamic framework, such as a square-gradient free energy functional, used to describe density fluctuations.

Combining \eqref{eq:force-mom} with \eqref{eq:lin-intforce}, we see that the forcing operator in the FDBE  \eqref{eq:fdbe-force} of a general D2Q$n$ model can be written as
\beqn K_{ab}(\kv) =
\left(
\begin{array}{cccc|ccc}
 0 & 0 & 0 & 0 &  &  &   \\
 \im k_x \left[c_s^2(\kv) - \cslb^2\right] & 0 & 0 & 0 &  & \b0 &  \\
 \im k_y \left[c_s^2(\kv) - \cslb^2\right] & 0 & 0 & 0 &  &  &  \\
\hline
  &  &  &  &  &  & \\
 \multicolumn{4}{c|}{\b0} &  & \b0 & \\
  &  &  &  &  &  &
\end{array}
\right)\,,
\eeqn
where the upper-left submatrix spans over the conserved modes. The generalization to the three-dimensional case is obvious.
Using the fact that the equilibrium distribution is of ideal gas form, $\bar m_a = \rho_0 \delta_{a1}$, the mode correlation matrix $\bG$ follows from eq.~\eqref{eq:g-correl} as
\beq
\bG(\kv) = \text{diag}\left[ S(\kv), \mu\rho_0 N_2, \ldots , \mu\rho_0 N_n \right]\,,
\label{eq:g-mat-forcing}
\eeq
with the structure factor being given by eq.~\eqref{eq:s-nonideal}.
From eq.~\eqref{eq:g-mat-forcing} we immediately see, that the mass parameter $\mu$ has to be chosen as
$$\mu = \frac{k_B T}{\cslb^2}\,,$$
in order to obtain the correct momentum correlations, $G_{22}=\ldots=G_{d+1,d+1}=\rho_0 k_B T$.
Thus, $\mu$ is identical in the ideal and the force-based non-ideal fluid model. This is not surprising, as both models employ the same equilibrium distribution.
With the time-evolution operator being given by
\beq L_{ab}(\kv)  = A_{ab}(\mathbf{k}) + \lambda_a \delta_{ab} - \hat K_{ab}(\kv)\,,\label{eq:L-nonideal}\eeq
the noise covariance matrix follows from the FDT \eqref{eq:dbe-fdt} after a bit of algebra as
\beq \Xi_{ab}(\kv) =  2\lambda_a G_{ab}(\kv) = 2\lambda_a \frac{\rho_0 k_B T}{\cslb^2} N_a \delta_{ab}\,,
\label{eq:noise-nonideal}
\eeq
where by construction $\lambda_a = 0$ for the conserved modes ($a=1,\ldots d+1$).

The above noise covariance is identical to the one obtained in the ideal gas model, eq.~\eqref{eq:noise-ideal}, implying that the non-ideal interactions have no effect on the dissipation.
This is physically expected as the forcing should represent a fully reversible contribution to the DBE dynamics.
Technically, this result depends crucially on the cancellation between terms originating from the interaction force by corresponding terms originating from the linear advection operator in the FDT. An fundamental prerequisite for these cancellations to occur is the assumption that the interaction force is of the general form given by eq.~\eqref{eq:lin-intforce}, with the speed of sound being related to the structure factor by \eqref{eq:s-nonideal}.
If these prerequisites are met, both contributions of forcing and advection to the FDT, eq.~\eqref{eq:dbe-fdt}, disappear from the final result, eq.~\eqref{eq:noise-nonideal}.
Conversely, the requirement to obtain a well-defined FDT obviously also allows one to impose constraints on the possible forms of the interaction force.

\subsubsection{Modified-equilibrium model}
\label{sec:yeo-dbe}
The Langevin extension of the modified-equilibrium model of Swift et al.\ \cite{swift_lattice_1995, swift_lattice_1996} has been analysed in \cite{gross_pre2010} based on the LBE dynamics. There, it was shown that relation \eqref{eq:f-correl-dbe} for the equal-time correlations of the distribution function had to be modified to obtain a well-defined noise covariance matrix. The latter turned out to be $k$-dependent and contained residual influences of the thermodynamic interaction model.
It will be informative to analyze the modified-equilibrium model starting from the DBE, and thereby compare it to the force-based model of the previous section. Where necessary, we will specify expressions in the D2Q9 basis given in appendix \ref{sec:basis-set}, for simplicity.

To derive the FDBE of the modified-equilibrium model, we note that in this case the equilibrium distribution $f\ueq$ is not just given by the projection of the full distribution onto the hydrodynamic subspace, $P^H_{ij} f_i$ [eq.~\eqref{eq:lin-feq}], as the second moment of $f\ueq$ is defined to reproduce an equilibrium non-ideal pressure tensor. The derivation leading to eq.~\eqref{eq:lin-DBE} can, however, easily adapted by noting that $\Lambda(f-f\ueq) = \Lambda(f-P^H f)+\Lambda(P^H f-f\ueq) = \Lambda^R f - \Lambda^R f\ueq$, since $P^H f\ueq = P^H f$ due to mass and momentum conservation.
Thus, in moment-space, we arrive at the FDBE
\beq
 \partial_t \delta m_a(\kv) + A_{ab}(\mathbf{k}) \delta m_b(\kv) =  - \lambda_a \delta_{ab} \left[\delta m_b(\kv) - \delta m_b\ueq(\kv)\right] + \hat \xi_a(\kv)\,,
 \label{eq:FDBE-yeo}
\eeq
where again, $\lambda_a=0$ for the conserved moments and $\delta m_a\ueq$ is the linearized equilibrium distribution in moment space. In this model, all non-ideal fluid interactions are represented by a pressure contribution to the equilibrium distribution. Specifically, for the D2Q9 basis set defined in appendix \ref{sec:basis-set}, the equilibrium moments follow as \cite{gross_pre2010}
\beq\delta m_a\ueq(\kv) = \{\delta \rho, \delta j_x, \delta j_y,
 d(\kv)\delta \rho, 0, 0, 0, 0, -d(\kv)\delta \rho\}\,,
\label{eq:meq-yeo}\eeq
where $d(\kv)\equiv  6\left[c_s^2(\kv) -\cslb^2\right]$ and $c_s(\kv)$ is a generalized speed of sound which is related to the structure factor $S(\kv)$ by the eq.~\eqref{eq:s-nonideal} above.
As seen in expression \eqref{eq:meq-yeo}, the non-ideal equilibrium contributions appear in a bulk pressure and in a ghost mode. Similar results are obtained also on three-dimensional lattices.
It was shown in \cite{gross_pre2010}, that the modified-equilibrium model requires the following equilibrium correlation matrix:
\beq
\bra \delta f_i(\kv) \delta f_j(-\kv) \ket = \frac{S(\kv)}{\rho_0} \bar f_i(\kv) \delta_{ij}\,.
\label{eq:f-correl-yeo}
\eeq
This is different from the expression eq.~\eqref{eq:f-correl-dbe} derived from continuum kinetic theory. The reason for these discrepancies can be traced back to the fact that, in contrast to conventional kinetic theory, the modified-equilibrium model employs a non-Maxwellian form of the equilibrium distribution.
With eq.~\eqref{eq:meq-yeo}, the time-evolution operator of the DBE \eqref{eq:FDBE-yeo} can be identified as
\beq L_{ab}(\kv) = A_{ab}(\kv) + \lambda_a \delta_{ab} - \lambda_4 d(\kv)\delta_{a4}\delta_{b1} + \lambda_9 d(\kv) \delta_{a9}\delta_{b1}\,.
\eeq
The noise covariance matrix then follows from the FDT \eqref{eq:dbe-fdt} after a bit of algebra as
\beq
\Xi(\kv) = \frac{2\rho_0 k_B T}{\cslb^2} \left(
\begin{array}{ccc|ccc|ccc}
 . & . & . & . & . & . & . & . & . \\
 . & . & . & . & . & . & . & . & . \\
 . & . & . & . & . & . & . & . & . \\
\hline
 . & . & . & N_4 \left[2-3c_s^2(\kv)\right] \lambda_e & . & . & . & . & 12 \left[c_s^2(\kv)-\cslb^2\right] (\lambda_e + \lambda_\epsilon) \\
 . & . & . & . & N_5 \lambda_s & . & . & . & . \\
 . & . & . & . & . & N_6 \lambda_s & . & . & . \\
\hline
 . & . & . & . & . & . & N_7\lambda_q & . & . \\
 . & . & . & . & . & . & . & N_8\lambda_q & . \\
 . & . & . & 12 \left[c_s^2(\kv)-\cslb^2\right] (\lambda_e + \lambda_\epsilon) & . & . & . & . & N_9\left[\frac{5}{4} - \frac{3}{4} c_s^2(\kv)\right]\lambda_\epsilon  \\
\end{array}
\right),
\label{eq:noise-yeo}
\eeq
where the dots indicate zeros for short.

While the noise obtained for the modified-equilibrium model exactly respects mass and momentum conservation, it is wavenumber-dependent and, hence, non-local, in contrast to the noise \eqref{eq:noise-nonideal} of the force-based model. These non-localities arise due to the modified equilibrium distribution employed in the model, c.f.\ eq~\eqref{eq:meq-yeo}, which is also responsible for the wavenumber-dependent bulk viscosity arising in the hydrodynamic equations \cite{gross_pre2010}.
Expression \eqref{eq:noise-yeo} is identical to the noise obtained from the LBE analysis of the modified-equilibrium model \cite{gross_pre2010}, except for lattice-induced contributions to the relaxation parameters (see below).
Similarly as for the ideal gas, we find that also for the modified-equilibrium model, the equilibrium correlation matrix and the advection operator fulfil the relation $\bA \bG = -\bG \bA^\dagger$. Hence, the contribution of the advection operator to the FDT identically vanishes, in contrast to the force-based model, where the advective contribution is canceled by a contribution from the forcing term.

\section{Fluctuating Lattice Boltzmann Equation}
\label{sec:flbe}
In order to transfer results of the previous section to the LBE, we apply a second-order accurate characteristics based integration to the FDBE, following \cite{nash_2008}.
The subsequent steps can be performed on the general (non-linear) FDBE
\beq
 \partial_t f_i + \mathbf{c}_i \cdot \nabla f_i = -  \Lambda_{ij} ( f_j - f_j\ueq ) + \Phi_i + \xi_i\,,
 \label{eq:fdbe}
\eeq
which we write in compact notation as
\beqn \partial_t f_i + \cv_i\cdot \nabla f_i = R_i(\rv,t), \eeqn
where $R_i = -\Lambda_{ij}(f_j - f_j\ueq) + \Phi_i + \xi_i$ is introduced for short.
Integrating over a time step $\Delta t$, we obtain
\beqn\begin{split}
f_i(\rv + \cv_i\Delta t, t+\Delta t) - f_i(\rv,t) &= \int_0^{\Delta t} ds R_i(\rv+\cv_i s,t+s) \\
&= \frac{\Delta t}{2}R_i(\rv+\cv_i\Delta t, t+\Delta t) - \frac{\Delta t}{2} R_i(\rv,t) + \Delta t R_i(\rv,t)\,.
\end{split}
\eeqn
Evaluation of the integral using the trapezium rule leads to a set of implicit finite-difference equations for the $f_i$,
which can be made explicit by introducing a set of auxiliary distribution functions
\beq f_i\uLB(\rv,t) \equiv f_i(\rv,t) - \frac{\Delta t}{2} R_i(\rv,t)\,. \label{eq:f-aux}\eeq
The time step is set henceforth to $\Delta t=1$ in lattice units.
Expressing $R_i$ in terms of the $f_i\uLB$, one obtains
\beq R_i=\left(\Imat+\frac{1}{2}\Lambda\right)^{-1}_{ij} [-\Lambda_{jk}(f_k\uLB - f_k\ueq) + \Phi_j + \xi_j]\,,
\label{eq:Ri}
\eeq
which leads to the fluctuating LBE (FLBE)
\beq f_i\uLB(\rv+\cv_i, t+1) = f_i\uLB(\rv,t) + \left(\Imat + \frac{1}{2} \Lambda\right)^{-1}_{ij} \left[-\Lambda_{jk}(f_k\uLB - f_k\ueq) + \Phi_j + \xi_j\right] \,.
\label{eq:lbe-gen}
\eeq
Note that the noise variables $\xi_i$ appear here just as another source term in addition to the forcing-term $\Phi_i$.
In order to write the LBE in moment space using an arbitrary basis set $T_a$, we compute
$(\Imat + \onehalf \Lambda)^{-1} = \left(\Imat + \onehalf T^{-1} \hat \Lambda T\right)^{-1} = T^{-1}\left(\Imat + \onehalf \hat \Lambda\right)^{-1} T = T^{-1}\left(\Imat - \onehalf \hat \Lambda\uLB\right) T$,
where in the last equation, we defined $\hat \Lambda\equiv \text{diag}[\lambda_a]$, $\hat \Lambda\uLB\equiv \text{diag}[\lambda_a\uLB]$ and introduced a set of collision parameters $\lambda_a\uLB$ by
\beq \lambda_a\ut{LB} = \frac{1}{\tau\ut{LB}} = \frac{1}{\tau_a + \onehalf} = \frac{2\lambda_a}{2+\lambda_a}\,.\eeq
These redefined collision parameters allow us to write the LBE in a form conventionally found in the literature.
With these definitions, we obtain $(\Imat + \onehalf \Lambda)^{-1} \Lambda = T^{-1} \text{diag}[2\lambda_a/(2+\lambda_a)] T = T^{-1} \hat \Lambda\ut{LB} T$, leading finally to the moment-space version of the FLBE \eqref{eq:lbe-gen},
\beq
f_i\uLB(\rv+\cv_i,t+1) = T_{ia}^{-1}\big[ m_a\uLB - \lambda_a\uLB (m_a\uLB - m_a\ueq) \\+ \left(1-\onehalf \lambda_a\uLB\right)m^F_a  + \left(1-\onehalf \lambda_a\uLB\right)\hat \xi_a \big]\,.
\label{eq:lbe-mom}
\eeq
We recognize in \eqref{eq:lbe-mom} the appearance of the well-known factor $\left(1-\onehalf \lambda_a\uLB\right)$ in front of the forcing-term \cite{laddVerberg_2001, guo_2002, nash_2008, guo_mrt_2008}, which is necessary to ensure a second-order accurate influence of the body force.
Since our derivation of the FLBE does not differentiate whether source terms originate from the random noise or a body force, the same factor naturally multiplies also $\hat \xi_a$.
Using eqs.~\eqref{eq:f-aux} and \eqref{eq:Ri}, the original DBE moments $m_a = T_{ai} f_i$ can be expressed in terms of the redefined LBE moments $m_a\uLB = T_{ai} f_i\uLB$ by
\beq m_a = m_a\uLB + \onehalf \left[-\lambda_a\uLB (m_a\uLB - m_a\ueq) + \left(1-\onehalf \lambda_a\uLB \right) (m^F_a + \hat \xi_a) \right]\,.
\label{eq:lbe-moments}
\eeq
In particular, the hydrodynamically relevant moments density, momentum and stress are obtained as
\begin{align}
\rho &= \sum_i f_i\uLB\,,\\
\rho u_\alpha &= f_i\uLB c_{i\alpha} + \onehalf F_\alpha\,,\\
S_{\alpha\beta} &= f_i\uLB Q_{i\alpha\beta} + \onehalf \left[-\lambda_a\uLB (f_i\uLB Q_{i\alpha\beta} - \rho u_\alpha u_\beta) + \left(1-\onehalf \lambda_a\uLB \right) (u_\alpha F_\beta + u_\beta F_\alpha + \xi_i Q_{i\alpha\beta})\right]\,,
\end{align}
where in the last equation, the expression for the second moment of the forcing term, eq.~\eqref{eq:forcing-term}, has been used. Note that due to eq.~\eqref{eq:noise-nonideal}, the noise gives a non-zero contribution to the instantaneous stress $S_{\alpha\beta}$.
The last two equations contain the well-known lattice corrections to the momentum and stress in force-based models \cite{laddVerberg_2001, guo_2002, nash_2008}.

The term $(1+\onehalf\lambda_a\uLB) \hat \xi_a\equiv \hat \xi_a\uLB$ in eq.~\eqref{eq:lbe-mom} defines the noise $\hat \xi_a\uLB$ appropriate to the FLBE in terms of the FDBE noise $\hat \xi_a$.
Both the ideal and the force-based non-ideal fluid FDBE considered above have identical noise covariances, given by eqs.~\eqref{eq:noise-ideal} and \eqref{eq:noise-nonideal}. Hence, in both cases the real-space covariance of the LBE noise follows as
\beq
\begin{split}
\bra \hat \xi_a\uLB(\rv) \hat \xi_b^{\text{LB}}(\rv')\ket &= \left(1-\onehalf \lambda_a\uLB\right)\left(1-\onehalf \lambda_b\uLB\right) \bra \hat \xi_a(\rv) \hat \xi_b(\rv')\ket = \frac{\rho_0 k_B T}{\cslb^2\, \Delta V} (2-\lambda_a\uLB)\lambda_a\uLB\, N_a \delta_{ab} \delta_{\rv,\rv'}\,.
\end{split}
\label{eq:noise-lbe}
\eeq
The factor $\Delta V$ arises from the lattice equivalent of the delta function and is taken as the volume of the elementary lattice cell. It reflects the fact that fluctuations become more pronounced with decreasing length scale.
Expression \eqref{eq:noise-lbe} is identical to the results previously derived by \cite{adhikari_fluct_2005, duenweg_statmechLB_2007} for the ideal gas LBE \footnote{Note that our variable $\mu$ corresponds to $m_p$ in ref.~\cite{duenweg_statmechLB_2007}}.
Comparing eq.~\eqref{eq:noise-lbe} to \eqref{eq:noise-ideal}, we identify the term $-(\lambda_a\uLB)^2$ as a lattice correction to the FDT of the FLBE, analogous to the well-known ``streaming contribution'' to the viscosity.

In contrast to the modified-equilibrium model, the derivation of the FDT for force-based non-ideal fluid models turned out to be technically cumbersome using the direct LBE approach introduced in \cite{gross_pre2010}. There, difficulties arise from the presence of the non-linear LBE advection operator, $A\uLB(\kv)_{ab}=T_{aj} \exp(-\im \kv\cdot \cv_j)T^{-1}_{jb}$, and the spatial discretization scheme of the interaction force. 
In contrast to the DBE, it appears to be technically demanding to construct a LBE forcing term that is fully reversible at the lattice level, while at the same time fulfil basic physical requirements (such as thermodynamic consistency \cite{wagner_consistency_2006, kikkinides_consistency_2008} or the absence of spurious currents \cite{wagner_spurious_2003}). 
In this regard, the DBE-based approach, as presented in this work, cannot fully replace a lattice treatment of the FDT \cite{dufty_lblangevin_1993, adhikari_fluct_2005, duenweg_statmechLB_2007, gross_pre2010}. Nevertheless, it provides a useful starting point for the construction of fluctuating LBEs.

\section{Simulations}
\label{sec:results}
We now investigate whether the fluctuating LBE of the force-based model derived in the preceding section can correctly reproduce some basic statistical mechanical results for non-ideal fluids.
Results for the ideal gas and the modified-equilibrium model have already been reported in \cite{adhikari_fluct_2005} and \cite{gross_pre2010}, where close agreement between simulation and theoretical expectations has been found.
Here, we first check whether thermal noise defined by eq.~\eqref{eq:noise-lbe} leads to the correct equilibration of all degrees of freedom in a LB simulation of a homogenous non-ideal fluid. Next, capillary fluctuations in a liquid-vapor interface are investigated as an important example for thermal fluctuations in an inhomogeneous system.
For the generic force-based model discussed in this work, spatially uncorrelated, ideal-gas-like noise with variance given by eq.~\eqref{eq:noise-lbe} is an exact consequence of the FDT for all wavenumbers. This form of noise is easily implemented in a simulation, as all noise modes $\xi_a$ can be drawn independently on each lattice site from a Gaussian distribution. For this purpose, the fast random number generator described in \cite{ladd_random_2009} can be successfully employed.

Simulations of the non-ideal fluid are performed using D2Q9 implementation of the model of Lee and Fischer \cite{lee_fischer_2006}, which is a refined variant of the generic force-based model of He, Shan and Doolen \cite{he_dbe_1998}.
The Lee-Fischer model is based on a square-gradient free energy functional, thus ensuring accurate reproduction of thermodynamics \cite{kikkinides_consistency_2008}.
It differs from the original He-Shan-Doolen model in that it employs a special discretization scheme for the derivative operators and uses an alternative, but thermodynamically equivalent way to compute the interaction force. Since the effect of the discretization generally becomes noticeable only at large wavenumbers, it is expected that both models are equivalent in the low-wavenumber region.

It must be remarked that the Lee-Fischer model shows a spurious drift (typically, an increase) of the total mass in a simulation box due to non-conservative discretization of the derivative operators \cite{chiappini_cpc2010, gross_stress_2011}. The magnitude of this effect is found to depend on the strength of the velocity gradients that exist in the system. If not properly handled, results for the structure factor can be spoiled seriously. A possible way to enforce mass conservation is to rescale all populations $f_i$ after each timestep by a global factor $x$ that is computed from the overall mass increase in the system, $x=\sum_{\bv{r}}\Delta m(\bv{r},t) / \sum_{\bv{r}}m(\bv{r},0)$. We found that this procedure gives best results for the structure factor.
Despite these insufficiencies of the Lee-Fischer model, its behavior under thermal noise is nevertheless worthwhile to investigate, as this model has the advantage of not being plagued by spurious momentum currents \cite{wagner_spurious_2003, lee_fischer_2006} and thus provides a promising candidate for the simulation of a range of complex fluid problems.

In our non-ideal fluid simulations, we employ a simple Landau double-well free energy density \cite{lee_fischer_2006, RowlinsonWidom_book}
\beq f_0(\rho) = \beta (\rho - \rho_V)^2(\rho-\rho_L)^2
\,,
\label{eq:f0-bulk}
\eeq
where $\rho_L$, $\rho_V$ are the prescribed equilibrium liquid and vapor densities and $\beta$ is a compressibility parameter.
The associated equation of state is given by $p_0 = \rho \partial_\rho f_0 - f_0$, from which the speed of sound follows as
$c_s^2 = \partial p_0 / \partial \rho = \rho \partial^2 f_0 / \partial\rho^2$.
Typically, instead of $\beta$, one rather prefers to specify the speed of sound, which is related to $\beta$ by
$$\beta = \frac{c_s^2(\rho_0)}{2\rho_0 (\rho_L-\rho_V)^2}\,,$$
where $\rho_0$ denotes the reference density for which $c_s$ was computed.
The shape of the equilibrium density profile assumes (far from the critical point) the mean-field form
\beq \rho\st{MF}(y) = \onehalf(\rho_L+\rho_V) + \onehalf(\rho_L-\rho_V)\tanh \frac{2y}{w}\,, \label{eq:denprof}\eeq
where the interface width $w$ is given by
\beq w = \sqrt{\frac{8\kappa}{\beta}} \frac{1}{\rho_L - \rho_V} = 4\frac{\sqrt{\rho_0\,\kappa}}{c_s}\,,
\label{eq:interf-width}\eeq
with $\kappa$ being the square-gradient parameter.
Finally, the surface tension can be expressed as
\beq \sigma = \frac{(\rho_L - \rho_V)^3}{6}\sqrt{2\kappa\beta} \,.
\label{eq:surf-tension}
\eeq

\subsection{Equilibration tests}
\begin{figure}[t]\centering
    (a)\includegraphics[width=0.32\linewidth]{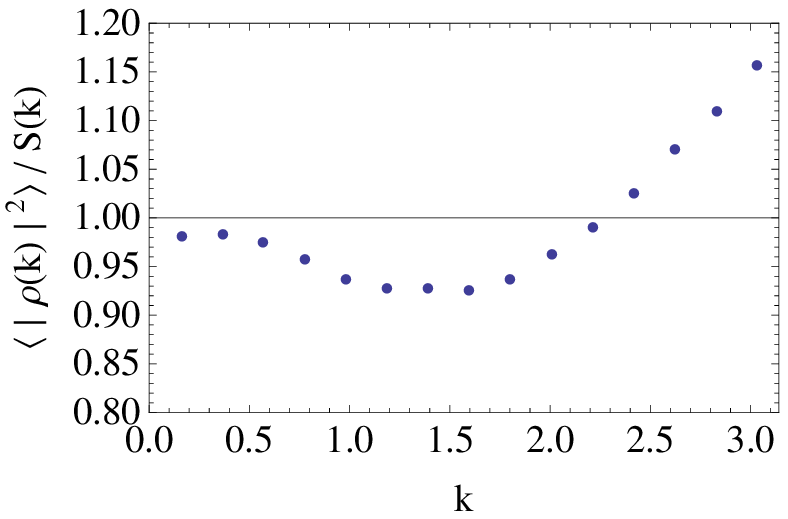}\quad
    (b)\includegraphics[width=0.23\linewidth]{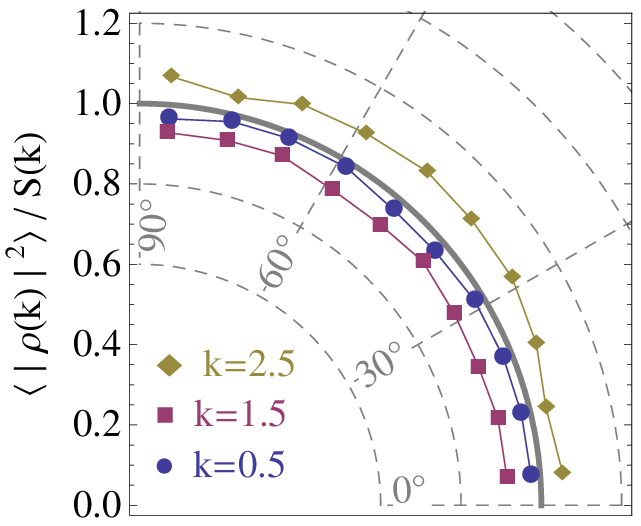}\quad
    (c)\includegraphics[width=0.32\linewidth]{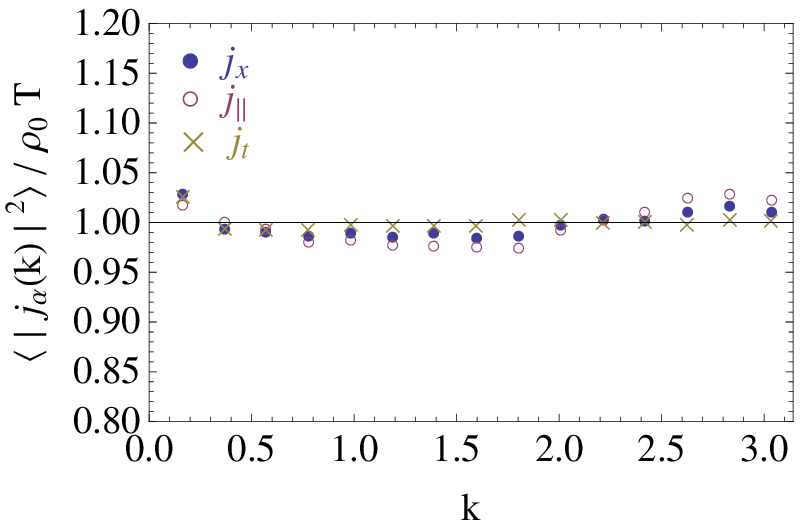}
    (d)\includegraphics[width=0.23\linewidth]{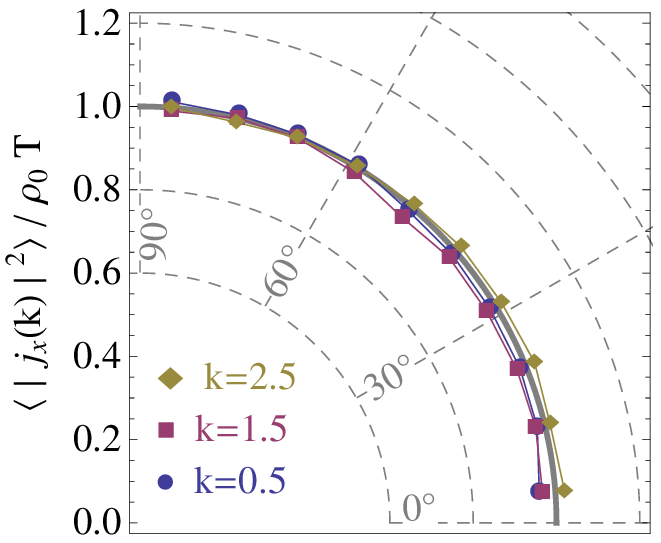}\quad
    (e)\includegraphics[width=0.32\linewidth]{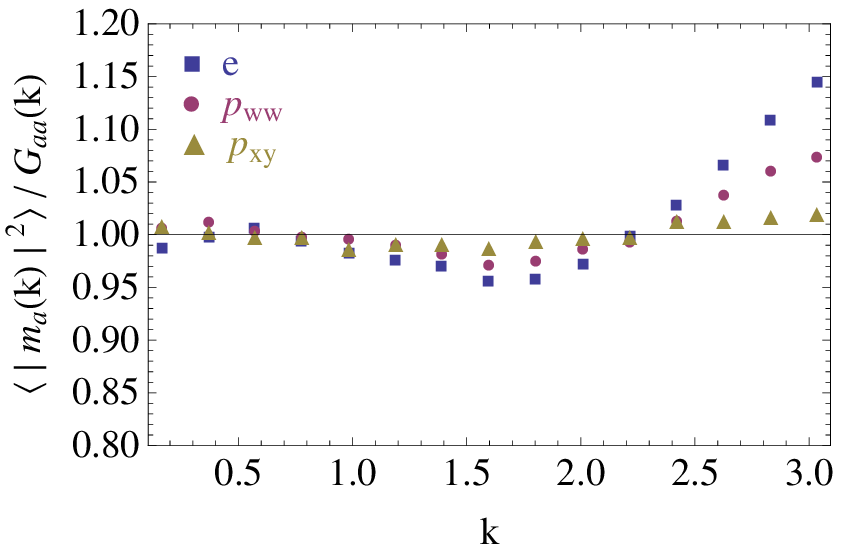}\quad
    (f)\includegraphics[width=0.33\linewidth]{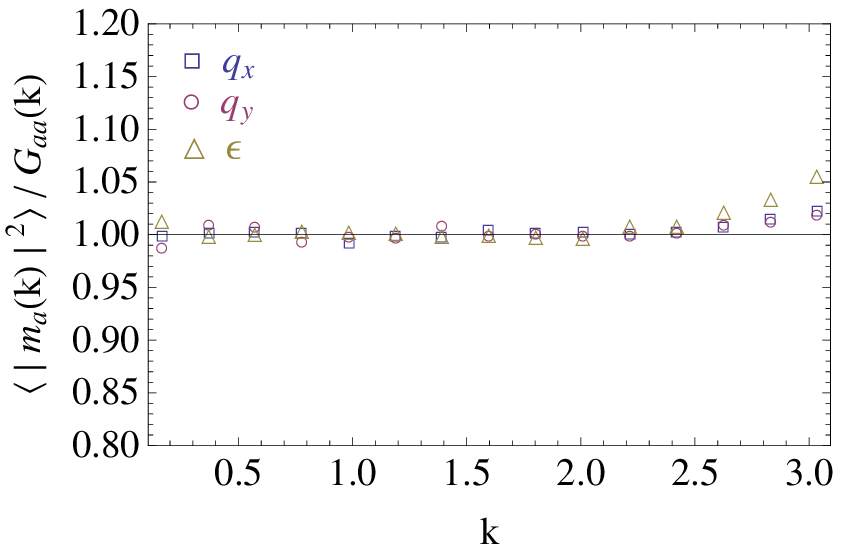}
   \caption{Thermal noise in the force-based model for $\kappa=0.08$, $c_s=0.265$, $\tau=1.0$. Equilibration ratios of (a,b) the density, (c,d) the momentum, (e) the transport and (f) the ghost modes. In (b) and (d) the dependence of the equilibration ratio of the density and momentum on $\theta$, when $\kv=(\cos\theta,\sin\theta)k$, is shown for several magnitudes of $k$. In the remaining plots, each data point represents an average over all directions in $k$-space at each magnitude $|k|$. $j_x$ denotes the $x$-component, $j_{||}$ the longitudinal and $j_t$ the transversal component (with respect to $\kv$) of the momentum $\bv{j}$.}
    \label{fig:force-unoise-fluct-kappa008}
\end{figure}

In a linear, homogenous fluid, Gaussian thermal noise described by eq.~\eqref{eq:noise-nonideal} is expected to produce Gaussian distributed fluctuations of the LB modes with a covariance matrix given by eq.~\eqref{eq:g-mat-forcing}.
We test this basic result by performing simulations in a uniform one-phase system of mean density $\rho_0=1.0$, choosing the square-gradient parameter and the speed of sound as $\kappa=0.08$ and $c_s=0.265$. The simulation box is a two-dimensional periodic domain of size $128\times 128$ lattice units (l.u.).
The fluctuation temperature is chosen as $T=10^{-7}$ (setting $k_B=1$ in l.u.), and all relaxation times are set to a value of $\tau = 1.0$.
Note that the stability constraint of LB, $|u| \ll \cslb$, together with relation \eqref{eq:momtm-correl-l} leads to an upper bound of $\cslb^2 \rho_0$ on the fluctuation temperature $T$ \cite{gross_pre2010}.
Simulation results are most conveniently compared to theoretical expressions by computing the equilibration ratio, which is defined as the ratio of the equal-time correlations $\bra |\delta m_a(\kv)|^2\ket$ of a LB mode divided by its expected variance $G_{aa}(\kv)$. This quantity is averaged over 400 simulation snapshots.
The equilibration ratio is computed for each wavenumber $\kv$ spanning the first quadrant of the first Brillouin cell of the reciprocal lattice. Here, a wavevector of $k_\alpha = \pi$ corresponds to a wavelength of 2 l.u.
The basis set $T_a$ used to obtain the modes $m_a$ from the populations $f_i$ is presented in appendix \ref{sec:basis-set}. For this choice, the modes are given by
$m_a=\{\rho, j_x, j_y, e, p_{ww}, p_{xy}, q_x, q_y, \epsilon\}$, where $e$ denotes a bulk stress mode, $p_{ww}$ and $p_{xy}$ are shear modes, and $q_x$, $q_y$ and $\epsilon$ are ghost modes.

Fig.~\ref{fig:force-unoise-fluct-kappa008} shows the equilibration ratios for all LB modes in the force-based model of Lee and Fischer. We observe that, while the momentum is equilibrated to high accuracy (Figs.\ \ref{fig:force-unoise-fluct-kappa008}c,d), the structure factor (Figs.\ \ref{fig:force-unoise-fluct-kappa008}a,b) shows an unusually large error for all but the smallest wavenumbers. The longitudinal momentum mode $j_{||}$ and the bulk stress mode $e$ resemble the error found in the structure factor to a certain extent, as is expected due to the coupling between these modes \cite{gross_pre2010}.
We have further noted a significant $\tau$-dependence of the density and momentum equilibration ratio at larger wavenumbers, prohibiting in fact the effective use of the Lee-Fischer model for values of $\tau$ much different from unity.
These effects are not present for the non-conserved modes, which are found to always remain well equilibrated up to $k\sim 2.5$.
Although a more detailed analysis of this model is out of the scope of the present work, one might attribute part of the observed discrepancies to the lack of microscopic reversibility of the forcing-term (see the discussion in sec.~\ref{sec:FDBE}). This is also indicated by the violation of global mass conservation caused by the particular spatial discretization scheme employed in the model.

\subsection{Capillary fluctuations}
The equilibration tests of the previous section are performed in a homogenous, and therefore, fully linear system. However, practical applications of non-ideal fluid simulations, including, for example, phase coexistence, usually entail the presence of nonlinearities.
Although the Langevin noise used in the present work was derived from linear equations, it is nevertheless expected to be applicable to small amplitude fluctuations around inhomogeneous equilibrium states (see the discussion in sec.~\ref{sec:FDBE}).
Crucially, then the local values of the density $\rho_0$ and the relaxation parameters $\lambda_a$ must be used in the computation of the noise variance \eqref{eq:noise-nonideal} on each lattice site.

As a basic example for such a fluctuating non-linear system we investigate in the following capillary fluctuations of a liquid-vapor interface \cite{mandelstam_1913, RowlinsonWidom_book}. We mention a number of previous simulations of fluctuating interfaces using lattice gas automata \cite{flekkoy_1995, flekkoy_1996} and a fluctuating ideal gas LBM \cite{rothman_2000}. Note that in the latter work, the interface was modeled as an elastic membrane, which is different from the present approach, where the interface is represented by a smoothly varying order parameter.

Capillary fluctuations are excited by the thermal noise in the bulk \cite{loudon_1980, grant_desai_1983} and can be described in terms of a local height function $h(\rv_{||})$, where $\rv_{||}$ denotes a position in the interfacial plane. In the present, two-dimensional situation, $r_{||} = x$, while we denote the perpendicular coordinate by $y$.
In the classical capillary wave theory \cite{buff_1965, evans_interface_1979}, $h$ is defined by the shift between the intrinsic density profile $\rho\st{int}$ and the instantaneous fluctuating density profile $\rho(\rv)$,
\beq \rho(\rv) = \rho\st{int}(y-h(\rv_{||}))\,.\label{eq:height-def} \eeq
In the harmonic approximation, the static spectrum of the local height fluctuations $h$ of a flat interface is given by
\beq \bra |h(\kv_{||})|^2 \ket = \frac{k_B T}{\sigma k^2}\,, \label{eq:capill-static}\eeq
where $\kv_{||}$ is the wavevector in the interfacial plane and $\sigma$ is the surface tension, eq.~\eqref{eq:surf-tension}.
For the present case of a Ginzburg-Landau model with bulk free energy given by eq.~\eqref{eq:f0-bulk}, far from the critical point, we can take for $\rho\st{int}$ the mean-field profile \eqref{eq:denprof} and obtain $h$ by fitting $\rho\st{int}$ to the density profile obtained from simulation.
This procedure smooths out (bulk-like) density fluctuations that are always present in the interface \cite{zittartz_1967, evans_1981, stecki_contrib_1998} and otherwise lead to deviations from the capillary wave theory predictions at high wavenumbers \cite{gross_pre2010}.

In order to test whether the static spectrum can be reproduced by our fluctuating non-ideal fluid model, simulations of a liquid stripe in a fully periodic, rectangular box are performed.
The extension of the stripe is taken as $1024\times 200$ l.u., and the box size accordingly as $1024 \times 400$ l.u.
Simulation parameters are $\rho_L=1.0$, $\rho_V=0.5$, $\beta=0.1$, $\kappa=0.08$, $T=10^{-7}$ and $\tau=1.0$.
The capillary spectrum is obtained, after neglecting an initial roughening period \cite{flekkoy_1995, flekkoy_1996}, by averaging over 2000 snapshots in a simulation running for $10^6$ timesteps, which is two orders of magnitude larger than the largest possible relaxation time of a capillary fluctuation in the system \cite{jeng_capvisc_1998}. 
Crucially, as we are working on a lattice, $k^2$ in eq.~\eqref{eq:capill-static} has to be replaced by the Fourier-transform of the proper one-dimensional discrete Laplacian, $2-2\cos k_{||}$. The difference between the continuum and lattice Laplacian becomes noticeable only for large wavenumbers ($k\gtrsim 1$).

\begin{figure}[t]\centering
     \includegraphics[width=0.4\linewidth]{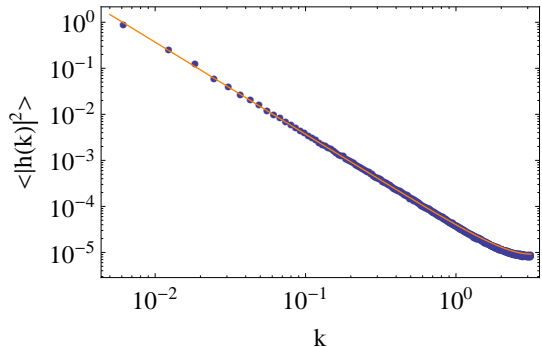}
   \caption{Capillary fluctuations of a planar one-dimensional interface in the fluctuating Lee-Fischer model. The equal-time spectrum of interfacial height fluctuations obtained from simulation [dots] is compared to the theoretical capillary structure factor [solid line, eq.~\eqref{eq:capill-static}]. $k$ denotes the wavenumber in the plane of the interface. Simulation parameters: $\rho_L=1.0$, $\rho_V=0.5$, $\beta=0.1$, $\kappa=0.08$, $\tau=1.0$, the interface width is approximately 5 l.u.}
    \label{fig:capill-static}
\end{figure}

Fig.~\ref{fig:capill-static} shows the static spectrum of the interfacial height fluctuations obtained for the fluctuating Lee-Fischer model.
We find perfect agreement between simulation results and the theoretical capillary structure factor \eqref{eq:capill-static} for practically all wavenumbers. We attribute the slight deviations for wavenumbers above $k\sim 1$ to the breakdown of the harmonic approximation on which eq.~\eqref{eq:capill-static} is based. These effects will have to be investigated in future works. We remark that a check of the variance of the velocity fluctuations parallel to the inhomogeneity indicated that the fluid remains well equilibrated within an error of a few percent.
We finally remark that essentially identical results can be obtained for capillary fluctuations in the modified-equilibrium model \cite{swift_lattice_1995} if definition \eqref{eq:height-def} is used for the determination of the interfacial height instead of a simple crossing criterion, as was employed in \cite{gross_pre2010}.

\section{Summary}
In this work, we have demonstrated how thermal noise can be incorporated into the discrete Boltzmann equation for the ideal and non-ideal fluid. The DBE is a precursor of the LBE, the latter being a well-established tool for fluid dynamical simulations. By linearizing the DBE and promoting it to a Langevin equation, fluctuations can be analyzed within the theory of non-equilibrium thermodynamics due to Onsager and Machlup.
The covariance of the Gaussian noise sources in the fluctuating DBE follow from a fluctuation-dissipation relation that is expressed in terms of the DBE time-evolution operator and the equilibrium correlations of the distribution function.
The equilibrium correlations are determined by invoking results of the kinetic theory of fluids.

Technically, the analysis of the DBE is simpler than of the LBE due to the linearity of the Fourier-transformed advection operator and the absence of lattice corrections that occur in various places of the LBE. We have shown how fluctuating LB models can be constructed in a straightforward way starting from the DBE: As the noise represents just another source term in the DBE (in addition to a possible body force) the noise covariance of the associated fluctuating LBE follows immediately using well-established relations between both equations.
Indeed, using the present approach to re-derive the noise covariance of the ideal gas model and the non-ideal fluid model of Swift et al. led to expressions in agreement with previous works based on the LBE \cite{adhikari_fluct_2005, gross_pre2010}.

As a central result of the present work, we applied the theory to the DBE of a non-ideal fluid model in which the thermodynamic interactions are mediated by a body force. We have outlined general requirements that such a model has to fulfil in terms of the structure factor and the expression of the interaction force in order to obtain a physically sound and mathematically well-defined noise covariance. In particular, the noise should neither be affected by forcing nor advection, as both terms represent reversible contributions to the dynamics. If these requirements are met, the noise in the LBE of a force-based non-ideal fluid is of the same form as in the ideal gas model.

Simulations of the fluctuating LBE version of the Lee-Fischer model indicated that satisfactory equilibration at all but the shortest length scales is indeed possible in this model with ideal-gas-type noise.
However, we also observed spurious influences of the relaxation time on the quality of the results. 
These findings are most likely the result of a not fully reversible forcing term of the Lee-Fischer model -- which also becomes manifest through the violation of mass conservation -- and have to be clarified in the future.

The further development of force-based FLBEs that overcome above mentioned deficiencies is now clearly put as a challenge for future works. In this regard, it would be desirable to test the present theory with other types of force-based models, such as two-distribution-function models derived from the original He-Shan-Doolen approach \cite{he_zhang_jcomp1999}.
Furthermore, it would be extremely interesting to assess the abilities of the FLBE under non-equilibrium conditions, such as a fluid under uniform shear \cite{machta_shear_1980, tremblay_noneq_1981}. Also, the extension of the presently available ``athermal'' FLBEs to include energy conservation remains an important objective for the future.

\section{Acknowledgments}
M.G.\ and F.V.\  acknowledge financial support by the Deutsche Forschungsgemeinschaft (DFG) under the Grant No.\ Va205/3-3 (within the Priority Program SPP1164) as well as funding from the industrial sponsors of ICAMS, the state of North-Rhine Westphalia and the European Commission in the framework of the European Regional Development Fund (ERDF).
M.E.C. acknowledges support from the Royal Society and funding from EPSRC Grant No.\ EP/E030173.

\appendix

\section{Non-ideal fluid model}
\label{sec:force-models}
We briefly review the force-based non-ideal fluid model of He, Shan and Doolen \cite{he_dbe_1998} and show that it fits the formal structure described in section~\ref{sec:force-nonideal}.
In the model of He-Shan-Doolen, the interaction force is derived from a Vlasov-type mean-field approximation taking into account an Enskog volume-exclusion effect. The resulting expression for the interaction force can be written as \cite{he_dbe_1998, he_zhang_jcomp1999}
\beq
\Fv\ut{int} = \kappa \rho \nabla \nabla^2 \rho - \nabla (p_0 - \rho \sigma_s^2)\,,
\label{eq:he-force}
\eeq
where $\kappa$ is a constant (``square-gradient parameter'') and $p_0$ is a non-ideal pressure defining the equation of state.
After Fourier transforming and writing $\delta p_0 = c_s^2 \delta \rho$, the linearized interaction force becomes
\beq \delta \Fv\ut{int} = \im \kv \left[c_s^2(\kv) -\cslb^2\right]\delta \rho\,, \eeq
where we have used the definition of the generalized speed of sound,
\beq c_s^2(\kv) = c_s^2 + \kappa \rho_0 k^2\,. \label{eq:he-cs2}\eeq
In order to compute the structure factor, we note that the interaction force \eqref{eq:he-force} can be derived from a square-gradient free energy functional of the form \cite{zou_multiphase_1999},
\beq
\mathcal{F}[\rho] = \int dV\left(f_0(\rho) + \frac{\kappa}{2}|\nabla \rho|^2 \right)\,,
\label{eq:freeE-funct}
\eeq
with $f_0$ being a bulk free energy that is related to the pressure by $p_0 = \rho \partial_\rho f_0 - f_0$.
The structure factor obtained from the linearized free energy functional has the usual Ornstein-Zernike form \cite{Chaikin_book, Hansen_ThoSL},
\beq S(\kv) = \frac{\rho_0 k_B T}{c_s^2(\kv)}\,,  \eeq
with $c_s^2(\kv)$ being the generalized speed of sound given by eq.~\eqref{eq:he-cs2}.

\section{Basis set}
\label{sec:basis-set}
\begin{table}[tb]
\begin{center}
\begin{tabular}{r | c | c | c | c | c}
$a$ & $T_{ai}$ & $N_a$ & $m_{a}$ & $m_a\ueq$ & $\lambda_a$ \\
\hline
 1 & 1                                    &  1   & $\rho$    & $\rho$ & 0\\
 2 & $c_{ix}$                             & 1/3  & $j_x$     & $\rho u_x$ & 0\\
 3 & $c_{iy}$                             & 1/3  & $j_y$     & $\rho u_y$ & 0 \\
 \hline
 4 & $3c_i^2-2$                           &  4   & $e$       & $3 \rho\left(u_x^2+u_y^2\right)$ & $\lambda_e$\\
 5 & $2c_{ix}^2-{c}_i^2$                  & 4/9  & $p_{ww}$  & $\rho \left(u_x^2-u_y^2\right)$ & $\lambda_s$ \\
 6 & $c_{ix} {c}_{iy}$                    & 1/9  & $p_{xy}$  & $\rho u_x u_y$ & $\lambda_s$ \\
 \hline
 7 & $(3c_i^2-4){c}_{ix}$                 & 2/3  & $q_x$     & 0 & $\lambda_q$\\
 8 & $(3c_i^2-4){c}_{iy}$                 & 2/3  & $q_y$     & 0 & $\lambda_q$\\
 9 & $9c_i^4-15{c}_i^2+2$                 & 16   & $\epsilon$  & 0 & $\lambda_\epsilon$\\
\hline
\end{tabular}
\end{center}
\caption{Basis set of the D2Q9 model used in the present simulations. $T_{ai}$ denotes the basis vector, $N_a$ its length, $m_{a}$ is the designation of the corresponding moment and $\lambda_a$ denotes its eigenvalue in the relaxation operator. $m_a\ueq=T_{ai} f_i\ueq$ is the expression for the corresponding moment of the ideal gas equilibrium distribution.}
\label{tab:modes-d2q9}
\end{table}

Table \ref{tab:modes-d2q9} shows the basis vectors $T_{ai}$ \cite{duenweg_statmechLB_2007} and the associated modes $m_a$ of the D2Q9 model used in the present simulations and in the theoretical analysis of the modified-equilibrium model (section \ref{sec:yeo-dbe}). Suitable basis sets in three dimensions can be found, for example, in \cite{dHumieres_MRT_2002, duenweg_statmechLB_2007}.
The parameters $\lambda_a$ denote the eigenvalues of $T_a$ under the relaxation operator $\Lambda$ of the DBE.
The first three rows cover the conserved hydrodynamic moments, the next three the non-conserved hydrodynamic moments, and the last three the ghost (or kinetic) moments.
The moment $e$ describes a bulk stress mode, which is related to the deviatoric stress $S_{\alpha\beta}$ defined in eq.~\eqref{eq:f-moments} by $e=3\tr\; S\,.$
The eigenvalue $\lambda_e$ of $e$ is related to the bulk viscosity $\nu_b$ by $\nu_b = \cslb^2/\lambda_e$.
The quantities $p_{ww} = S_{xx} - S_{yy}$ and $p_{xy} = (S_{xy} + S_{yx})/2$ are shear modes, with a common eigenvalue $\lambda_s$ related to the shear viscosity $\nu_s$ by $\nu_s = \cslb^2/\lambda_s$.
The ghost sector finally consists of a ghost density mode $\epsilon$ and ghost vector current $q_\alpha$, with eigenvalues $\lambda_\epsilon$ and $\lambda_q$, respectively.
Using the numerical expressions for the lattice velocities, the transformation matrix can be written in compact form as
\beq T = \left(
\begin{array}{ccccccccc}
 1 & 1 & 1 & 1 & 1 & 1 & 1 & 1 & 1 \\
 0 & 1 & 0 & -1 & 0 & 1 & -1 & -1 & 1 \\
 0 & 0 & 1 & 0 & -1 & 1 & 1 & -1 & -1 \\
 -2 & 1 & 1 & 1 & 1 & 4 & 4 & 4 & 4 \\
 0 & 1 & -1 & 1 & -1 & 0 & 0 & 0 & 0 \\
 0 & 0 & 0 & 0 & 0 & 1 & -1 & 1 & -1 \\
 0 & -1 & 0 & 1 & 0 & 2 & -2 & -2 & 2 \\
 0 & 0 & -1 & 0 & 1 & 2 & 2 & -2 & -2 \\
 2 & -4 & -4 & -4 & -4 & 8 & 8 & 8 & 8
\end{array}
\right)
\,.
\label{eq:TD2Q9}
\eeq

\section{Advection operator}
\label{sec:adv}
It is instructive to write down the explicit expression for the DBE advection operator in moment space, $A_{ab}(\kv) = -\im T_{aj} \bv{k}\cdot \bv{c}_j T_{jb}^{-1}$. Taking the basis set $T_a$ defined by eq.~\eqref{eq:TD2Q9}, we obtain
\beq
A_{ab}(\kv) = -\im\left(
\begin{array}{ccc|ccc|ccc}
 . & k_x & k_y & . & . & . & . & . & . \\
 \frac{k_x}{3} & . & . & \frac{k_x}{6} & \frac{k_x}{2} & k_y & . & . & . \\
 \frac{k_y}{3} & . & . & \frac{k_y}{6} & -\frac{k_y}{2} & k_x & . & . & . \\
\hline
 . & 2 k_x & 2 k_y & . & . & . & k_x & k_y & . \\
 . & \frac{2 k_x}{3} & -\frac{2 k_y}{3} & . & . & . & -\frac{k_x}{3} & \frac{k_y}{3} & . \\
 . & \frac{k_y}{3} & \frac{k_x}{3} & . & . & . & \frac{k_y}{3} & \frac{k_x}{3} & . \\
\hline
 . & . & . & \frac{k_x}{6} & -\frac{k_x}{2} & 2 k_y & . & . & \frac{k_x}{6} \\
 . & . & . & \frac{k_y}{6} & \frac{k_y}{2} & 2 k_x & . & . & \frac{k_y}{6} \\
 . & . & . & . & . & . & 4 k_x & 4 k_y & .
\end{array}
\right)\,.
\label{eq:adv-op}
\eeq
This is identical to the $O(k)$-term in the expansion of the LBE-advection operator \cite{gross_pre2010},
\beq A\uLB_{ab}(\kv)\equiv T_{aj} \exp(-\im \kv\cdot \cv_j)T^{-1}_{jb} = \Imat + A_{ab}(\kv) + O(k^2)\,.
\label{eq:lbe-adv}
\eeq


\end{document}